\documentclass[manuscript, review=False, screen, acmsmall]{acmart}

\renewcommand\footnotetextcopyrightpermission[1]{} 
\usepackage{longtable}
\AtBeginDocument{%
  \providecommand\BibTeX{{%
    \normalfont B\kern-0.5em{\scshape i\kern-0.25em b}\kern-0.8em\TeX}}}

\setcopyright{acmcopyright}
\copyrightyear{2019}
\acmYear{2019}
\acmDOI{10.1145/1122445.1122456}

\usepackage[ruled]{algorithm2e}
\usepackage[colorinlistoftodos,prependcaption,textsize=tiny]{todonotes}
\usepackage[justification=centering]{caption}
\usepackage{csquotes}

\usepackage[utf8]{inputenc}

\usepackage{booktabs}
\usepackage{refcount}
\usepackage{footnote}
\usepackage{pdflscape}
\usepackage{rotating}
\usepackage{longtable}
\usepackage{afterpage}

\newcommand{\young}[1]{{\color{black} #1}}

\newcommand{\yd}[1]{{\color{black} #1}}

\newcommand{\tb}[1]{{\color{black} #1}}

\makesavenoteenv{tabular}
\makesavenoteenv{table}

\SetAlFnt{\small}
\SetAlCapFnt{\small}
\SetAlCapNameFnt{\small}
\SetAlCapHSkip{0pt}
\IncMargin{-\parindent}

\AtBeginDocument{%
  \providecommand\BibTeX{{%
    \normalfont B\kern-0.5em{\scshape i\kern-0.25em b}\kern-0.8em\TeX}}}

\setcopyright{acmcopyright}
\copyrightyear{2019}
\acmYear{2019}
\acmDOI{10.1145/1122445.1122456}

\begin{document}

\title{A Survey on Computational Politics}


\author{Ehsan ul Haq}
\email{euhaq@connect.ust.hk}
\affiliation{%
  \institution{Hong Kong University of Science and Technology}
  \streetaddress{Clear Water Bay, Kowloon}
  \country{HKSAR}
}

\author{Tristan Braud}
\email{braudt@ust.hk}
\affiliation{%
  \institution{Hong Kong University of Science and Technology}
  \streetaddress{Clear Water Bay, Kowloon}
  \country{HKSAR}
}

\author{Young D. Kwon}
\email{ydkwon@cse.ust.hk}
\affiliation{%
  \institution{Hong Kong University of Science and Technology}
  \streetaddress{Clear Water Bay, Kowloon}
  \country{HKSAR}
}
\author{Pan Hui}
\email{panhui@cse.ust.hk}
\affiliation{%
  \institution{Hong Kong University of Science and Technology}
  \streetaddress{Clear Water Bay, Kowloon}
  \country{HKSAR}
}
\affiliation{%
  \institution{University of Helsinki}
  \streetaddress{}
  \city{Helisnki}
  \country{Finland}
}

\renewcommand{\shortauthors}{Haq et al.}
\begin{abstract}

Computational Politics is the study of computational methods to analyze and moderate users' behaviors related to political activities such as election campaign persuasion, political affiliation, and opinion mining. With the rapid development and ease of access to the Internet, Information Communication Technologies (ICT) have given rise to massive numbers of users joining online communities and the digitization of analogous data such as political debates. These communities and digitized data contain both explicit and latent information about users and their behaviors related to politics. For researchers, it is essential to utilize data from these sources to develop and design systems that not only provide solutions to computational politics but also help other businesses, such as marketers, to increase users’ participation and interactions. In this survey, we attempt to categorize main areas in computational politics and summarize the prominent studies in one place to better understand computational politics across different and multidimensional platforms. e.g.,  online social networks, online forums, and political debates. We then conclude this study by highlighting future research directions, opportunities, and challenges.

\end{abstract}

%
%




%
%






\fancyfoot{}

\maketitle

\thispagestyle{empty}

\section{Introduction}\label{sec_intro}


The Internet and digitization of data, particularly in recent years, have changed the way we process information.  The inception of web 2.0 opened the gateway for online communities. Data generation, collection, and transformation  are happening at an exponentially increasing rate, following the rapid growth of websites focusing on people-to-people interactions and content sharing.
This new generation of websites has paved the way for not only gathering massive amounts of data on their users but also conveying personalized messages to a very diverse audience. Following this trend, political parties, influence groups, and individuals have rapidly seized these new communication media to propagate their ideas.

\textbf{Computational Politics. } 
The definition of computational politics is \enquote{applying computational methods to large datasets derived from online and offline data sources for conducting outreach, persuasion, and mobilization in the service of electing, furthering or opposing a candidate, policy or legislation}~\cite{tufekci_engineering_2014}. These computational methods vary from statistical analysis, probabilistic models, and visualization of data to cover the different socio-political behaviors of users.
By devising models, frameworks, and systems, computational methods allow \yd{us} to discover users’ \yd{socio-political} behaviors and analyze how information propagates within communities. The resulting insights can then be exploited for political purposes such as opinion mining, polls, marketing, and political campaigns.

\textbf{Historical Context. }
From a historical perspective, the usage of data for political motives is not a new thing. However, the methods and goals evolved dramatically. Traditionally, data collection happened at an aggregated level and used in the context of broadcasting mediums for political campaigns. Political messages were then targeted to the aggregate audience through broadcast media. Such audiences consisted typically of a group of people from the same locality, consumers of similar products, viewers of a similar program, or people around the same age group. 

\tb{Computational politics traditionally} used to focus on political persuasion and marketing, 
\yd{where mass media played an essential role in shaping public opinion. On the other hand,} social scientists suggested a more holistic approach to shaping and spreading public opinion. \yd{For instance,} in 1947, E. Bernay published \emph{The Engineering of consent}~\cite{bernays_engineering_1947} that provides an engineering approach to getting people to support ideas and programs. Among other things, it stresses the importance of engineering the whole action sequence in parallel for people to transform their thinking and the message to be conveyed. 

Over time, media has evolved into different forms. With the appearance of social media, it became the source of data instead of just spreading messages. However, the importance of the medium to convey messages, as expressed by Mcluhan in 1964~\cite{mcluhan1964understanding}:  \enquote{Medium is the message}, is still relevant in many situations today. Different type\yd{s} of platforms and data sources offer \yd{highly} diverse interaction styles to the users. 
H. Farell~\cite{farrell2012consequences} discusses the influence of the Internet on politics.
Due to the prominent usage of the Internet in everyday political life, the author suggests that it will be increasingly difficult to study a specific field without analyzing its relationship with the Internet. As such, he suggests a shift in political science towards integrating the Internet and the underlying data as an indissociable part of the field. 
Vinogradova et al.~\cite{vinogradova_political_2017} consider the development of mass communication tools as part of evolution. Traditional political messages fit into the new templates with the apparition of new tools that utilize the Internet. These templates give additional information in a new interface while coexisting with older information. However, there is still a gap between current political tools, i.e., mass communication, and new communication tools such as social media. Bridging this gap would allow \yd{us} to integrate and process information to draw a nuanced, multi-faceted, and multi-dimensional picture of political events. \textit{Data activism} is a concept introduced by Milan and Gutirrez~\cite{milan_citizens_2015} in a study of the new communication medium for political purposes. Data activism refers to citizens seizing the potential of big data for social change, at the intersection of data analysis, journalism, activism, and citizens and media empowerment. Data activism can be considered as a novel, decentralized, and data-based form of citizen media, redefining the relationship between citizens and data.

\young{\textbf{Highlights of This Survey.}}
There is an increasing number of research work focusing on computational politics. \textcolor{black}{However, to the best of our knowledge, no prior work compiles the related work at one place in a systematic approach.}
With this survey, we try to fill this gap by categorizing the existing literature on computational politics
into five major areas:
\young{\textit{(1) Community and User Modeling, (2) Information Flow, (3) Political Discourse, (4) Election Campaigns, and (5) System Design}}. We further study each category through the prism of its major related research areas.
For each of these areas, we report on the related work covering data from different kind\yd{s} of online communities such as social networks, crowd-sourced forums, and user-generated content sharing communities. Besides, we describe some works based on the digitized versions of political debates during elections and in the legislative assemblies of different nations. We report on these works following two descriptive frameworks, namely User and Data, allowing us to analyze the data features from two different perspectives. These frameworks enable us to highlight the similarities and differences between various data sources.

Since computational politics is a topic at the intersection of different fields such as political science, government policy designs, and computer science, it has been studied from multiple aspects, resulting in thousands of studies. In this survey, we decide to focus on the perspective of computer science and computational methods. We therefore primarily cover the literature coming from two libraries: ACM Digital Library and IEEE Xplore, which are well-established references for studies related to computer and data science. However, we also consider the major related studies from prominent libraries in other fields.

The rest of the paper is organized as follows.

\young{
\begin{itemize}
\item Section \ref{sec:framework} presents the categories and research areas in computational politics, along with the key concepts used in these areas. Also, we explain descriptive frameworks featuring data sources and widely used computational methodologies.
\item Section \ref{sec:survey} surveys recent works, in which we follow the categorization defined in Section~\ref{sec:framework}. We also report the methodology used along with the data and the results or findings of the computational methods. 
\item Section \ref{sec:discussion} identifies future research directions and recommendations based on the limitations of the surveyed works in Section \ref{sec:survey}.
\end{itemize}
}


\begin{figure}[t]
        \centering
        \includegraphics[width=
\textwidth, keepaspectratio]{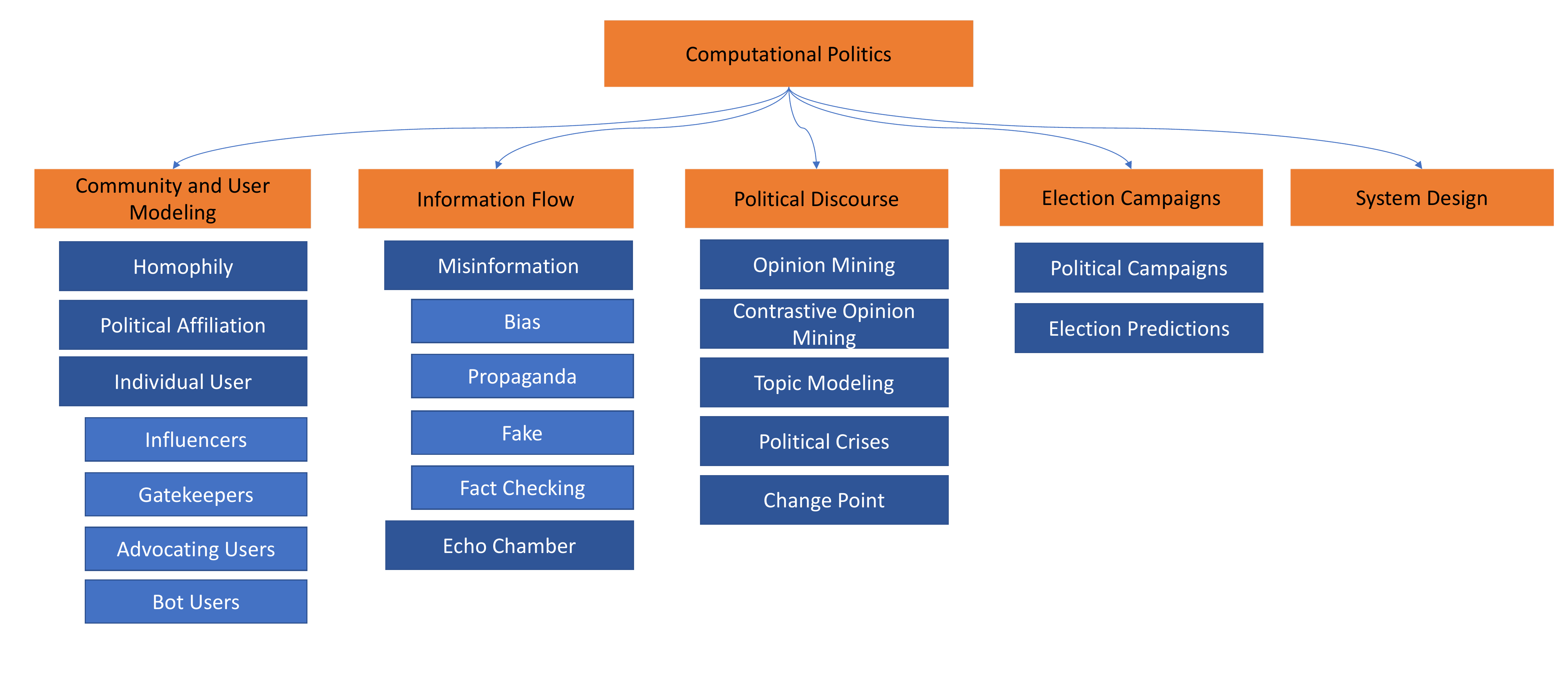}
        \caption{Categorization of Computational Politics }
        \label{fig:categorization}
    \end{figure}%

\section{Categorization and Descriptive Framework} \label{sec:framework}

In this section, we first categorize the computational politics based on the studies we report in this survey. We distinguish five major categories and then further classify them into major research areas as shown in Figure~\ref{fig:categorization}. We then present the two major descriptive frameworks used in computational politics.
\tb{Finally, we present the common computational methodologies used in research.}

\subsection{Categorization of Computational Politics} \label{sec:classification}

Analyzing the major works on the computational aspects of politics lead us to identify five major areas: Community and User Modeling, Information Flow, Political Discourse, Election Campaigns, and System Design studies. These categories give a high level hierarchy of the work covering different aspects. The first category (Community and User Modeling) deals with the approaches that focus on the users while the second (Information Flow) focuses on the information and data propagation in the respective media. The third category (Political Discourse), although closely related to the others, is one of the well studied topic in digital humanities and social media analysis, and in general look at the collective behavior of the population set. The fourth category (Election Campaigns) cover the work related to an important event in democratic process such as elections. The last category (System Design) studies the system developed to compliment the studies in the prior areas. Overall this classification approach covers user, data, discourse and systems.  We define these categories as follows.

\begin{itemize}
\item \textbf{Community and User Modeling:} Work in this domain focuses on modeling the online behavior of individual users and communities. Most of the work is related to the analysis of homophily, influencer users, gatekeepers, and detecting the political alignments of users. Many works in the category are closely related to information propagation in the networks, as such users play a critical role in the way information is propagated within a network.
\subitem {
 \textit{ Homophily}: The term homophily refers to the phenomenon where people with similar interests form a link with each other as in the proverb \textit{``birds of a feather flock together''}~\cite{mcpherson2001birds}. We discuss the work related to homophily, e.g., finding homogeneous communities, in Section~\ref{sec:survey}.
}
\subitem{ \textit{Political Affiliation}: Political affiliation refers to the partisan behavior of users where they tend to support or show positive sentiment for some political entity. Such results can further be used to understand the community and to predict the voting tendencies of users.}

There is a number of work that focuses on the individual users from certain perspectives such as advocating and influencing users for some campaigns. Use of bots for marketing and propaganda campaigns is also well known and several works focus on such areas. 
\subitem {\textit{Advocating users}: Advocating users refers to users that actively participate in communities to influence other users. We identify two main categories of advocating users: 
 \textit{Influencers}: In social network analysis, influencers are users whose content has the largest reach and are the most susceptible to affect the behavior of other people. Such users are key to efficient marketing and opinion engineering. 
\subitem{ \textit{Gatekeepers}: Gatekeeping\footnote{\url{https://en.wikipedia.org/wiki/Gatekeeping_(communication)}} is the process of information filtration during information propagation. In online communities, individuals act as gatekeepers between different cross-opinionated communities or in the information flow paths between users. Identification and behavior analysis of such users are important for studying data diffusion paths. 
}}

\item \textbf{Information Flow:} These approaches deal with the information and its propagation in the network. We classify this in two major areas: Misinformation and Echo Chambers. 
\subitem{\textit{Misinformation} further specializes in the detection of bias and fake news along with fact-checking. Another mostly discussed area in this domain is the analysis of echo chambers.}
\subitem{\textit{Echo Chamber}: In social media and online forums, the echo chamber refers to a situation in which communication and repetition result in the strengthening of opinions inside some community. Such communities are usually homogeneous, and such an effect can result in the creation of bubble effects and also cause potential bias in the media. }

\item \textbf{Political Discourse:} Political discourse deals with a broader range of topics related to politics and shows an overall analysis of demographics and communities together with information dissemination. Further hierarchies, in this section, include opinion mining, topic modeling, detecting political crises or events and change points, i.e., a study of factors that lead to changes in opinion. These can be used to find the topics that people are discussing at a given time and their opinions about those topics. Such analysis can be used to help users to effectively engage with each other~\cite{johnson2000civil}.   
\item \textbf{Election Campaigns:} 
Work in this area is related to effective engagement of the online audience, doing large scale opinions polls, running and managing political campaigns. We identify two main fields of study. 
\subitem{\textit{Political Campaigns}: Social media has been widely used for organizing political campaigns and increase the political engagement of users.}
\subitem{\textit{Election Prediction}: Election Predictions are crucial for many organizations, ranging from political parties to mass media. Online media allows for a broader range of opinions and demographics than traditional surveys and permits, in turn, to refine the predictions.}

\item \textbf{System Design:} This Section includes studies that propose a new system design related to any of the research areas mentioned above.

\end{itemize}

\subsection{Descriptive Frameworks}

We discuss multiple kinds of data-based studies in this work. It is therefore important to look at the sources of data, distinguish their particular features, and relate them to the corresponding approaches. 
In general, we can classify data sources into two categories: \young{(1) Social Networks and Content Sharing Sites, and (2) Political Debates}. These data sources can be looked upon from two perspectives based on the research focus: the \textit{User} perspective and the \textit{Data} perspective. In this section, we describe both perspectives and their main attributes. 

\subsubsection{Data Perspective}\label{sec:data_pers}
Throughout this document, we highlight the importance of data in computational politics. However, the lack of standard and centralized data sources makes data very heterogeneous. Most of the computational methods rely on data collected from social networks like Facebook and Twitter. In this section, we describe the core features of these heterogeneous data sources.  

\begin{figure}[t]
        \centering
        \includegraphics[width=
\textwidth, keepaspectratio]{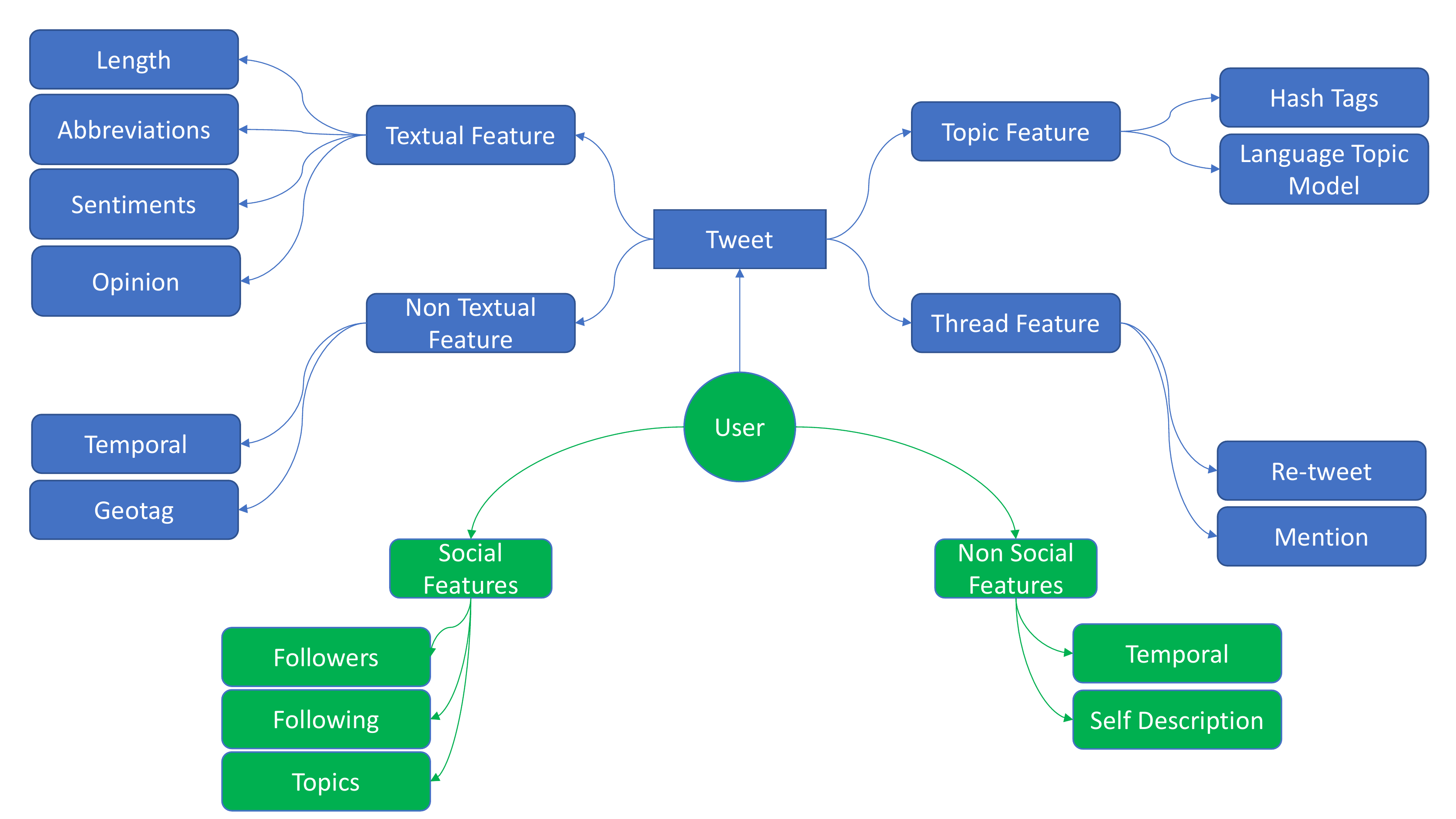}
        \caption{Entities and Attributes in Twitter  }
        \label{fig:Twitter_pers}
    \end{figure}

Ellison et al.\cite{ellison2007social} define social network sites as \enquote{Web-based services that allow individuals to (1) construct a public or semi-public profile within a bounded system, (2) articulate a list of other users with whom they share a connection, and (3) view and traverse their list of connections and those made by others within the system.}
Both Twitter and Facebook match these three properties. However, representation and interfaces are very different from one another. We describe Twitter’s framework in Figure~\ref{fig:Twitter_pers}. Users share messages called Tweets with their network. These tweets present four kinds of attributes.

\begin{itemize}
\item \textit{Textual Features}: Textual features include length, sentiments, writing style, and message conveyed. Twitter has a restriction of 280 characters (formerly 140), so many users try to use shortened versions of words. This usage significantly changes the methods of natural language processing.   
\item \textit{ Non-Textual Features}: Non-textual features compose the tweet's metadata, and capture the temporal and spatial information of the tweet. 
\item \textit{Thread Features}: Thread features highlight the information flow in the network. A Retweet is a tweet shared by a user other than its original author. A Mention is the direct use of a user's Twitter handle on another user's tweet.
\item \textit{Topic Features}: Restriction of the length results in using hashtags to highlight what the tweet is about. Hashtags are potential candidates to find topics in tweets along with language topic models. 

\end{itemize}

 Facebook presents similar features, with the exception of a few attributes. Facebook messages are usually referred to as posts and do not have a length restriction. Non-textual features of a Facebook post carries additional emotional reaction information. Thread features use the term \textit{tag} instead of \textit{mention}, and users can share posts from each other. Hashtags can still be found but not as implicitly as in Twitter.

\begin{figure}[t]
        \centering
        \includegraphics[width=
\textwidth, keepaspectratio]{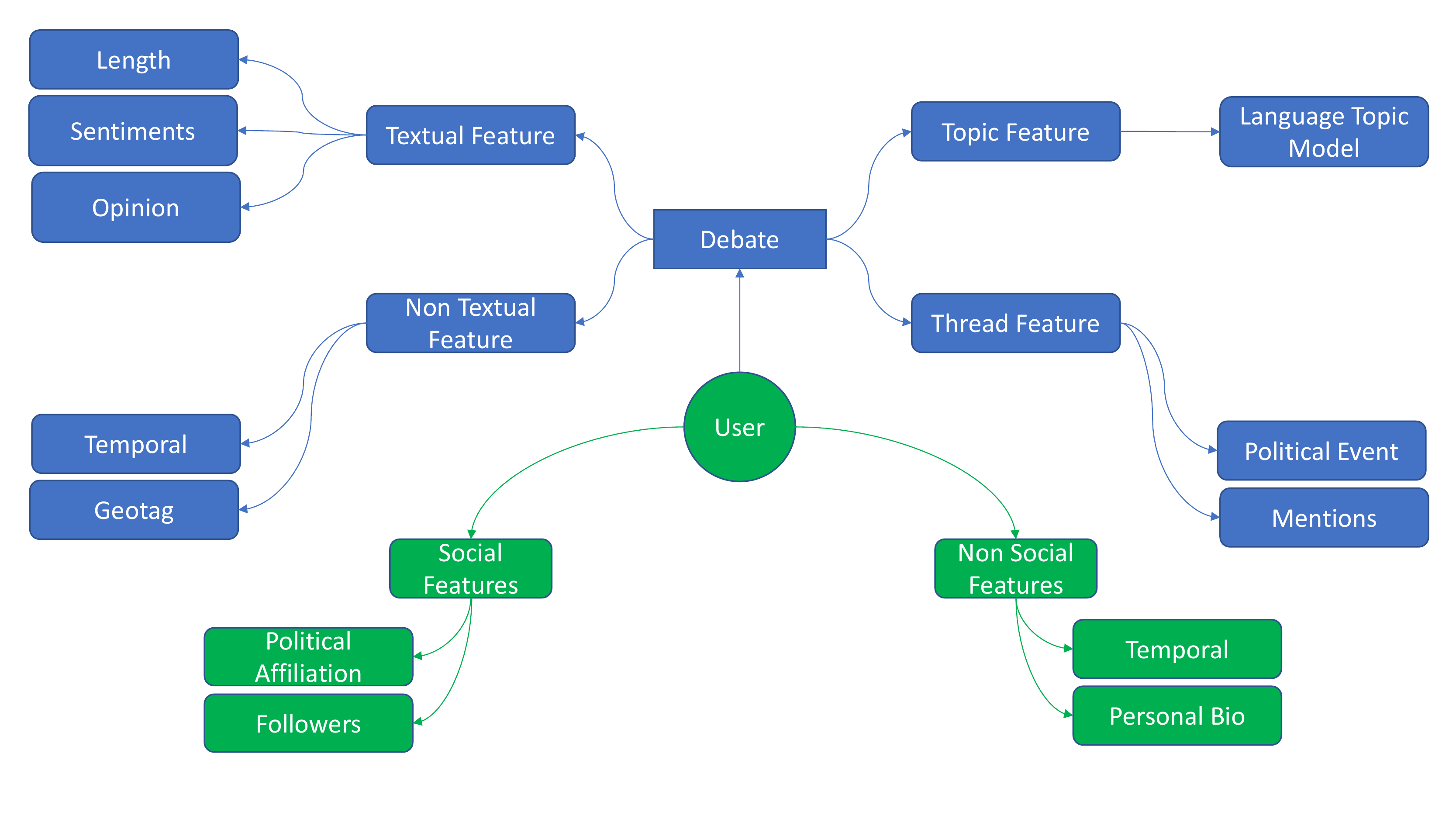}
        \caption{Entities and Attributes in Political Debates}
        \label{fig:debate_pers}
    \end{figure}

The other major data source in computational politics is political debates transcripts. Such a medium presents less latent information as compared to general users on social networking sites. Figure~\ref{fig:debate_pers} describes the features related to political debates. There are four kind of attributes: Textual, Non-Textual, Topic, and Thread Features. Textual, Non-Textual, and Topic Features remain the same as for social networks. However, the length is not a constraint as debates are generally quite long. 
Thread Features are different in political debates. In this particular case, thread feature is context-dependent as political debates are either in continuation of earlier political debates in elections or legislative assemblies sessions such as European Parliament or US election debates. Mentions in Political debates correspond to the references to other politicians and political events during the speech.

\subsubsection{User Perspective} \label{sec:user_pers}
Users are the main source of data regardless of the communication medium and form of the networks. Many latent attributes can be derived from users descriptions analysis. 
Social and Non-Social features of users, as shown in Figure \ref{fig:Twitter_pers} and \ref{fig:debate_pers}, show the community of users through their connection network with other users and entities such as communities or affiliations. 
Users’ affiliations can range from explicit to latent. For instance, in the case of social media, some users may mention their political affiliation. Others may choose not to mention it explicitly while showing a strong preference in their actions. Apart from being members of the same communities, users may also connect with each other in follower/following (friends) relationships. Some works combine users and data feature together to get other forms of connections between users. For instance, Retweet networks represent connections that do not necessarily follow social connections on Twitter.

\subsection{Computational Methodologies}

\begin{table}[t!]
  \begin{center}
    \begin{tabular}{*5c}
        \toprule
        Media& Network&Linguistic&Hybrid\\
        \toprule

        Twitter& *& *&*\\
        \hline
        Facebook& *& *&*\\
        \hline

        Blogs/Website& *& *&*\\
        \hline

        Political Debates& & *&\\
        \bottomrule
        
    \end{tabular}

  \end{center}

  \caption{Approaches used on different data sources/mediums}
  \label{table_medium_approaches}

\end{table}

Computational methods used in the research studies are varied from machine learning to linguistic processing,  to visualization and analysis tools. We categorize these methods into three approaches: \textit{Network Based, Linguistic Based, and Hybrid}. The network-based approach looks at the data in the form of the network. Social features of users such as follow relations in Twitter and thread features of content such as replies to a user are considered to build a network graph in which users form the nodes of the network and an edge between them can show the relation (such as follow, mention, reply, retweet, like, etc.) between two users. Network properties such as edge weights, centrality, or nodes distance are used in computational methods~\cite{kwon_self-disclosure_asonam19}. Examples include Follower/Following Network and Retweet Network. Purely linguistic-based approaches are applied primarily to the textual features of data. This approach is used for tasks such as sentimental analysis. Finally, some hybrid approaches not only consider the network but also include the textual and other content (e.g., temporal features) into the analysis of users' behaviors in the community. Table~\ref{table_medium_approaches} shows the medium and the nature of the computational approach that has been used in different papers. 
To achieve these approaches, most studies use one or several of the following methods.

\begin{itemize}
\item Classification Methods: Classification methods assign unknown observation to any of the identified categories. Machine learning algorithms~\cite{kotsiantis2007supervised} are common for binary or multi-label classification in tasks including political affiliation, fake news, bias detection, and political discourse analysis. Several methods based on supervised, unsupervised, and semi-supervised learning have been used. The most common algorithms are binary classification such as Support Vector Machines (SVM) and a probabilistic classifier such Naive Bayes. 
\item Term Frequency-Inverse Document Frequency (TF-IDF): TF-IDF methods~\cite{jones2004statistical} are based on textual content. TF-IDF is one of the most commonly used methods in text analysis and extraction of political topics, opinions, and political discourse analysis. TF-IDF is the product of two terms: Term Frequency and Inverse Document Frequency. It focuses on the frequency of certain keywords in the text and the importance of that keyword in all of the documents. 
\item Latent Dirichlet Allocation (LDA): LDA~\cite{blei2003latent} is based on a generative process that links the text in the document to the topics. In computational politics, it is used for topic extractions from the content (e.g., tweets). It allows us to find the topics in which people are interested.
\item Exploratory Studies: Exploratory studies focus on the analysis of data based on the visualization of linguistic features or network features. Some studies also perform empirical analysis on community networks~\cite{larsson2012studying}.

\end{itemize}

\section{Digital Century} \label{sec:survey}
With the evolution of computing,  computational methods have become paramount to analyzing and predict the political behaviors of individuals on the Internet. 
In this section, we report on these works according to the classifications defined in Section~\ref{sec:classification}.

\subsection{Community and User Modeling}
In computational politics, a number of questions revolve around the users’ behavior, their interactions, and the evolution of communities over time. These interactions, together with generated data, allow us to interpret users actions and to classify individuals and groups into specific categories. We first consider studies related to homophily. Homophily refers to similar people grouping together and is one of the most intuitive ways of classifying users. We then move to another critical latent attribute of users: their political affiliation. Finally, we focus on advocating and influencer users, a  category of individuals who are particularly active on social media.

\subsubsection{Homophily} \label{subsect: homophily}
As described in Section~\ref{sec:framework}, homophily represents the tendency of individuals to form a community based on similarity of opinions. As such, this concept can help describe the behavior of users as a community. In particular, homophily can explain the reinforcement of topics, as well as some information diffusion behaviors that may ultimately lead to the creation of echo chambers.

Grevet et al.~\cite{grevet2014managing} conducted interviews and surveys to find out how diverging opinions can affect social network engagement and friendship. Surveys are based on politically active users. The study reveals that connected users with less homophily are more likely to break connections due to political differences. This phenomenon happens even though social network sites are beyond the physical boundaries of communities and allow users with different demographic categorization to interact with each other. A study by Bastos et al.~\cite{bastos2013tweeting} analyzes the community structures on the usage of hashtags and linguistic content. The authors analyze the graph network constructed by the Clique Percolation Method (CPM) and other clustering algorithms~\cite{blondel2008fast,rosvall2008maps}. The study finds, from a Twitter dataset, that communities are formed based on linguistics and topics. Overlapping users in different hashtags are clustered around similar linguistic terms. However, activism-related hashtags and topics escape these linguistic barriers. Garimella et al.~\cite{garimella2018political} analyze echo chambers and user behavior as bipartisan and gatekeepers, and studies users’ polarity for content consumption and production. The paper identifies gatekeepers as users with bipolar spectrum exposure. The authors analyze the network and measure the retweet/favorite rates and volumes to predict partisan and gatekeeper users. The study shows that a user who tries to break the bubble effect of echo chambers has to pay the price in terms of the under-appreciation of their content.

A study by Fraisier et al.~\cite{fraisier2017uncovering} aims at detecting communities of like-minded users on Twitter from the network structure. The study uses the standard community detection algorithms including Modularity Maximisation, Information diffusion, and Random walk on a graph network based on retweet and mentions. Though limited in generalizations the study finds that the retweet network plays a much more vital role for such tasks than the mention  network\footnotemark\footnotetext{’ mention’ is the term used when a Twitter user refers to some other Twitter user in tweet}. Another study~\cite{ozer2016community} uses Heider’s Structural network to develop the endorsement network from the tweets and develops three non-negative matrix factorization frameworks. 
This study uses the non-negative matrix factorization method to find the political communities in the Twitter network. Both the content and endorsement network give complimentary accuracy to political community detection. Their algorithm achieves a purity of .897 on the Twitter dataset of UK users.

There are several models which allow users to link and unlink from other users.  A study by Perl et al.~\cite{perl2015Twitter} predicts link formation and dissolution with several models for a politician based on a dataset of German Politicians \footnotemark\footnotetext{GESIS Dataset: \url{https://gesis.org}}. The study considers seven models: Similarity, Popularity, Reciprocity, Social Embeddedness, Attraction, and Support. The paper shows that politician link formation is strongly affected by the collective link formation by the party and its members. 
The paper~\cite{borge-holthoefer_content_2015} studies the dynamic behavior of Egyptian Twitter users switching sides due to the civil law and order situation. The authors train a multi-class SVMLight classifier to predict the polarity of tweets to be either pro, anti, or neutral to the military. The classifier was trained on data  collected from Twitter\footnotemark\footnotetext{data available: \url{http://alt.qcri.org/~wmagdy/EgyMI.htm}}. This paper also develops weighted network models of retweets on a three day aggregated period. The study displays the political sentiments of the tweets. Pro-Islamist and anti-military users became louder as compared to secularist and pro-military users after the Egyptian military coup in 2013. \yd{Xu et al.~\cite{xu_structures_2013} study the unfollow behaviors of users on Twitter. The authors use an actor-oriented modeling (SIENA) to investigate the effects of homophily, status, reciprocity, embeddedness, and informativeness on tie dissolution.}

Romero et al.~\cite{romero_differences_2011} compare the information diffusion in terms of hashtags on Twitter. Political hashtags are different from other hashtags related to Music, Movies, Celebrities, Sports, Games, Idioms, and Technology. The authors build the network graph based on the \textit{mention} network between users if a user has mentioned another user at least three times. The consistency and persistence measures are defined, and approximation curves are used to study the adoption of the hashtag by a user. 
A user is considered to be exposed to a hashtag if an edge links him to any other user who has used the hashtag. This work also shows that controversial political hashtags are persistent, and this plays a significant role in adoption. 
Another study by Shi et al.~\cite{shi_cultural_2017} uses Twitter data from US users to study the community division that goes beyond the partisan division. This study is based on the co-follow network on Twitter, that is users who follow both political and cultural entities. The methodology uses the following statistical measures: political alignment index, political relevance index, and political polarization index using kurtosis methods. This research shows that a user’s choice of cultural entities (music, movies, restaurants) is profoundly affected by partisan alignments.  

A study~\cite{thuermer_internet_2016} highlights the effect of Internet tools on people participation in political decision-making processes based on German political parties through interviews. It shows that the Internet significantly affects the internal operations of the parties. Alternatively, the Internet can have a higher exclusion effect on people based on Internet access and usage patterns. However, such cases are scenario and context-dependent. 

Millennials are the people who will form the most significant portion of the electorate worldwide, estimated around 36.4 percent by 2020. Douglas et al.~\cite{douglas_politics_2014} survey the young demographics of Facebook users and show that social media plays an essential role in a voter’s perception of candidates because of the information that is curated by their social network on Facebook. This information is presented in their feed, even though most of those users do not explicitly search for it. The paper also shows that a candidate’s interaction with the rest of the community impacts users’ perception. Another study by Wang and Mark.~\cite{wang_engaging_2017} finds that young Facebook users display an imbalance between new consumption and participation. Such users might restrict themselves from participation to avoid conflict. These works show that both the user’s offline network and online network matter for user’s political efficacy, that is the trust they put in political entities.

Use of social media has become a primary part of many people's lives, particularly to share and observe important events. Gorkovenko and Taylor~\cite{gorkovenko_politics_2016} study the use of social media by people while watching the American Political Debates. Using social media, in addition to the TV broadcast, gives users the freedom to express their opinion, listen, and understand other’s opinions at a larger scale.  
A study by Margetts~\cite{margetts_understanding_2016} uses the terms “micro-donation” in terms of time spent by people for “tiny-acts” of different actions such as sharing, liking, commenting, and many other platform-specific actions. The actions lead to massive chain reactions that can create large-scale disruptions in the information flow and the creation of pluralism. That phenomenon is highly unpredictable and dynamic in nature. However, it can still generate effects like the Arab Spring protests where social media platforms were used to convey and amplify messages and actions.

\begin{table}
	
    \caption{Work in Homophilly }

    \begin{tabular}{p{0.30\linewidth}p{0.15\linewidth}p{0.15\linewidth}p{0.30\linewidth}}
      \toprule
      \multicolumn{1}{c}{\textbf{Paper}} & 		\multicolumn{2}{c}{\textbf{Feature\footnote{`*' shows the feature used in method. No `*' sign in any of the columns shows that work is a qualitative study based on survey or interviews} }} 
      &\multicolumn{1}{c}{\textbf{Method}} \\
  \toprule
  
   & \textbf{Linguistics} & \textbf{ Network}  &\\
\cmidrule{1-4}

Grevent et al. 2014 \cite{grevet2014managing}  & & &LDA, GBDT, Surveys, Interview\\
\hline
Bastos et al. 2013 \cite{bastos2013tweeting} &  *& *&  Clustering, Clique Percolation Method\\
\hline
Garimella et al. 2018\cite{garimella2018political} &&*&Polarity Measure, Network Parameters\\
\hline
Fraisier et al. 2017 \cite{fraisier2017uncovering} &&*& Standard Community Detection Algorithms\\
\hline
Ozer et al. 2016\cite{ozer2016community} &*&&Heider's structural balance theory, Nonnegative matrix factorization  \\
\hline
Perl et al. 2015 \cite{perl2015Twitter} &*&  &Link Prediction, Social Connection Forming models \\
\hline
Borge-Holthoefer et al. 2015 \cite{borge-holthoefer_content_2015} &*&&SVM Light Multiclass classifier  \\
\hline
Romero et al. 2011 \cite{romero_differences_2011} &*&*& Exposure Curve, Hashtag diffusion network analysis  \\
\hline
Shi et al. 2017 \cite{shi_cultural_2017} &*&*& Co- follower network analysis, empirical study \\
\hline
Thuermer et al. 2016\cite{thuermer_internet_2016} &*&*& Empirical study, Interviews \\
\hline
Douglas et al. 2014 \cite{douglas_politics_2014} &*&*&  Observational, Interviews  \\
\hline
Yiran Wang and Gloria Mark. 2017\cite{wang_engaging_2017} &*& &Observational, Interviews \\
\hline

Katerina Gorkovenko and Nick Taylor. 2016 \cite{gorkovenko_politics_2016} && &Observational, Interviews \\
\hline

Helen Margetts. 2016 \cite{margetts_understanding_2016} && &Observational \\
\hline
\yd{Xu et al. 2013 \cite{xu_structures_2013}} &  & * & Observational \\

\bottomrule

  \end{tabular}
  \label{t_db_topics_stats}

\end{table}

\subsubsection{Political Affiliation}

Political affiliation identification focuses on assigning a political class label based on the partisan inclination of users. Predicting political affiliation is a prominent topic in computational politics. Most online users do not mention their political affiliations explicitly. However, their interactions with other users and systems can be used to figure out this information.

Pennacchiotti and Popescu~\cite{pennacchiotti_machine_2011} in 2011 introduced one of the first works using a machine learning framework for political classification. This framework relies on the Gradient Boosted Decision Trees (GBDT) learning algorithm~\cite{friedman2001greedy}.  It derives four feature classes from the user's Twitter profile: profile, tweeting behavior, linguistic features of a tweet and user's network features based on followers and followed users. It shows from these classes of features that content features of the tweet are highly related to the political orientation of the user. Large scale topic modeling is also consistent and reliable. To observe the content of tweets, this study uses LDA and sentimental analysis to find the topics discussed along with the hashtags and sentimental words used as features. Domain-specific words are much more important for better results compared to all keywords. Social characteristics are also significant features because users tend to interact with certain celebrities or famous people. Rao et al.~\cite{rao_classifying_2010} use the stacked-SVM-based approach to find the latent attributes of users based on the linguistic features of their tweet. To extract features, the authors apply lexical and sociolinguistics models on tweets. The analysis shows that features showing possessiveness are highly correlated with political affiliation even though there is no political content in the tweets. 
Another study~\cite{chiu_predicting_2018} uses linguistic features from Facebook data to train classifiers and predicts if the political tendencies of social media posts are right-wing or left-wing. Models are trained on sentiments and word frequency in the text. The authors build the Term Document Matrix (TDM) based on Term Frequency (TF) and the TF-IDF score of the words from the posts. The authors report that TDM with TF is very large and the terms with sparse occurrences are removed to reduce the sparsity.
TF-IDF helps in focusing on terms that are important in different posts. Opinion lexicons from SentiWord Net~\cite{kamps2004using} and The Opinion Lexicon~\cite{hu2004mining} are used for sentiment analysis of the posts. Naïve Bayes, k-Nearest Neighbor (kNN), SVM, AdaBoost, C4.5 Decision Tree, and Classification and Regression Trees (CART) are used in the Weka tool \footnote{\url{https://www.cs.waikato.ac.nz/ml/weka/}}. In general, a left-wing prediction is easier than right-wing because of the higher number of words.  The study also shows that sentiment analysis effectiveness depends on the choice of algorithms. TF-IDF using decision tree produces an F1 score of up to .95 and is better than using TF. However, there is a difference in classification for each of the lexicons, 1NN and Naive Bayes perform better on the opinion lexicon and the SentiWordNet lexicons, respectively.
A study by Thomas et al.~\cite{thomas_get_2006} uses the transcripts of the debates from the presidential election to find support or opposition to the proposed legislation. The authors train SVM models for classification based on the features extracted from the political transcripts. The proposed method takes into consideration the speaker argument on some topics and whether there is any agreement between two different speakers on the same topic. These features are then added for same speaker and different speaker argument constraints. These debates may be divided into several segments, as speakers may give their opinions or answers to other speakers' arguments during the speech. The study concludes that models considering the relationship between the segments of speech perform better. 

Online articles may contain a few sentences in the quotes that represent the author's stance and contains topical information. A study by Awadallah et al.~\cite{awadallah_polaricq:_2012} introduces a classification method to map the quotation from online sources into a fine-grained political topic hierarchy along with the analysis of the content polarity. Unary and Binary features are extracted from the text in the form of topics and sentiment-topic. These are further expanded with antonym and synonym relations according to the sentiment. Classification models are trained to predict polarity using language models with n-gram, unary, binary, and unary features and achieve a precision of 74 percent. Bouillot et al.~\cite{bouillot_french_2012} introduce two variants of TF-IDF called TF-IDF adaptive and TF-IDF adaptive-normalized to find topics from political communities. This paper also shows that these new variants can be used to find community membership for users. The study used a Twitter dataset from the French presidential elections in 2012. The proposed method achieves a precision of  0.97.  Takikawa and Nagayoshi~\cite{takikawa_political_2017} study echo chambers and polarization based on political ideology among Japanese Twitter users. The study focuses only on the users who have a reciprocal follow relationship on Twitter. The authors used the Louvain method for community detection and LDA methods for topic modeling. Their study conforms to other studies and shows that political ideology is prominent in the Japanese Twitter space, and different echo chambers discuss non-overlapping topics. 
A study by Castro et al.~\cite{castro_back_2017} shows a methodology to predict the political alignment at the state level during the Venezuelan elections. The linguistic features in a tweet can represent the ideology of a geographic region, and hence can be used to predict the alignment of the region. The study uses the TF-IDF method and log-normalized TF-IDF method, which respectively achieve an accuracy of 79.17 and 87.50 percent.
Social networks provide a different kind of interactions that are platform-specific such as like, share, retweet, and the like. However, these interactions sometimes do not reflect the negative sentiments in user interactions. Users can also form negative links based on disliking or disagreement among users. Ozer et al.~\cite{ozer2017negative} use political datasets from three different countries (The UK, The US, and Canada) consisting of tweets for general elections and Brexit to predict such negative links through unsupervised learning, with up to 93.3 percent accuracy.  This study also shows that the prediction of such a link helps in detecting communities in polarized settings such as political affiliations.

Golbeck and Hansen~\cite{golbeck_computing_2011} introduce a three-step process to find the political inclination of users as liberal or conservative. The authors use a scale ranging from 0 to 1, 0 being most conservative and 1 being most liberal. These scores are obtained from the Americans for Democratic Actions (ADA) \footnotemark\footnotetext{\url{http://www.adaction.org/}}. The average of these scores is assigned to followers of labeled users and then mapped to target users. A paper by Makazhanov and Rafiei~\cite{makazhanov_predicting_2013} finds the political affiliation of  Twitter users based on their interactions with political entities.
The study uses the distant supervised learning approach to build predictive models for political affiliations. Language models (LMs) for each political party are built based on the tweets of political candidates. They perform term-weighting using Kullback-Leibler (KL) divergence on unigram and bigram terms from tweets. The LMs are used to create the interaction profiles of political parties. Predictive models are trained on 51 features. Logistic Regression and J48 classifiers are used and show different results, in terms of Precision, Recall, and F score, for different political parties based on the tweeting and interaction behavior of political parties' followers. In a temporal analysis, the study shows that the political affiliation of users change as a campaign-related event happens.  
Another study~\cite{conover_predicting_2011} predicts the political affiliation of a Twitter user based on the linguistic features of his tweets and the properties of their communication network. A communication network is modeled as two undirected weighted graphs: retweet and mentions. The weight of an edge between two users is based on the number of retweets and mentions between them.
SVM classification models are trained on linguistic features. Three different experiments using TF-IDF, Hashtags, and Latent Semantic Analysis~\cite{deerwester1990indexing} of hashtags are performed to evaluate the best feature set. The study also performs label propagation on the communication network. Training models on hashtags results in better accuracy than the models trained on a full text. On the other hand, community properties such as mention and retweet networks increase the accuracy in affiliation detection to 94 percent as compared to 91 percent when only hashtags are used. 
Castro and Vaca~\cite{castro_national_2017} use Twitter data for predicting users' political alignment and finding a political course in the national landscape during the Venezuelan election campaigns. The proposed scheme is based on building a political dictionary using topic modeling with LDA and assigning term scores. Louvain modeling is used on political hashtags to build the clusters. Linear SVM algorithms are used to predict political alignment. Additional dimensional reduction on dictionary data improves the result and also uses political tweet volume. The proposed method forecasts the election results with an accuracy of 98.7 percent.

\begin{table}

\begin{center}
\caption{Work in Political Affiliation Domain}

\begin{tabular}
{p{0.25\linewidth}p{0.15\linewidth}p{0.15\linewidth}p{0.10\linewidth}p{0.20\linewidth}}
\toprule
\multicolumn{1}{c}{\textbf{Paper}} & \multicolumn{2}{c}{\textbf{Feature\footnote{`*' shows the feature used in method. No `*' sign in any of the columns shows that work is a qualitative study based on survey or interviews} }} &\multicolumn{1}{c}{\textbf{Media}} &\multicolumn{1}{c}{\textbf{Accuracy}} \\
\toprule

  & \textbf{Linguistics} & \textbf{ Network} & &\\
\cmidrule{1-5}
Pennacchiotti and Popescu. 2011\cite{pennacchiotti_machine_2011} & *& *& Twitter& Decision Tree\\
\hline
Rao et al. 2010\cite{rao_classifying_2010}& *& & &Stacked SVM\\
\hline
 Chiu and Hsu. 2018\cite{chiu_predicting_2018} &*&&Facebook& TDM, TF-IDF, Classification Algorithms \\
\hline
Thomas et al. 2006\cite{thomas_get_2006} &*&&Debates&SVM \\
\hline
Awadallah et al. 2012\cite{awadallah_polaricq:_2012} &*&&Political Debates \footnote{\url{www.debatepedia.com}}& Classification Algorithms \\
\hline
Bouillot et al. 2012 \cite{bouillot_french_2012} &*&&Twitter & TF-IDF \\
\hline
Takikawa and Nagayoshi. 2017\cite{takikawa_political_2017}&*&*&Twitter &Community Detection (Louvain)/LDA \\
\hline
Castro et al. 2017\cite{castro_back_2017} &*&&Twitter &TF-IDF \\
\hline
Golbeck and Hansen. 2011\cite{golbeck_computing_2011} &&*&&Political Preference Scoring\\
\hline
Makazhanov and Rafiei. 2013\cite{makazhanov_predicting_2013} &&*&Twitter& Distant Supervised Learning  \footnotemark\footnotetext{Results varies for different parties}\\
\hline

Conover et al. 2011\cite{conover_predicting_2011} &*&*&Twitter & SVM, TF-IDF \\
\hline

Castro and Vaca. 2017\cite{castro_national_2017} &*&*&Twitter & TF-IDF,PCA, SVM \\

\bottomrule

\end{tabular}
\label{t_db_topics_stats}
\end{center}
\end{table}

\subsubsection{Advocating/influencer Users}
In previous sections, we highlighted how data from diverse platforms could be used for predicting community-level phenomenons. However, not only are users the main source of data, they are also a critical factor in its diffusion. Therefore, it is critical to analyze individual users based on their role. In social networks, influencer or advocating users are users who affect the rest of the community through information propagation and opinion changes. Thus, it is important to recognize these users for marketing and campaign activities. 

A study by Ranganath et al.~\cite{ranganath2016understanding} categorizes the methods used by Twitter users for political campaigns. They analyze the content of messages to ultimately finds advocating users. The study focuses on the persuasion and propagation strategy employed by such users. For persuasion, the authors measure the emotions and stress along with the topics detected by LDA~\cite{blei2003latent}. Advocating users target some users (popular users) more than others. Moreover, they propagate each other's messages to their respective communities and hubs. Based on such propagation strategies, the authors present a framework, using tensors, to classify the user either as an advocate or an ordinary user. 
Jürgens et al.~\cite{jurgens2011small} show that certain users in a small-world network act in a distinguishable way to curate and diffuse information to the rest of the network. Such users are referred to as the New Gatekeepers and are a significant source of political information diffusion. The authors use the centrality entropy of the network as a measure and label the vertices as gatekeepers. Removing such gatekeepers from the network can reduce the central entropy.
In another study, Hemsley et al.~\cite{hemsley2017call}, analyze the behavior of political candidates based on their messages and the retweet from the public. The authors focus on the role of gatekeepers in information diffusion. They perform content analysis using SVM classification to classify the tweets into different themes: Advocacy’, ‘Informative’, ‘Call to Action’, ‘Attack’, ‘Conversational’, ‘Ceremonial’. Multilevel regression analysis is used to perform information diffusion analysis. The study shows that middle-level gatekeeper (users with more than twice the followers of general users) are more active in the network's information flow. The authors therefore suggest considering these users as influential users. They also relate their work to another study~\cite{katz19552006} that suggests that opinion leaders are more influential as compared to media elites because opinion leaders transform the information taken from media elites to serve their purposes and diffuse it to the masses. 

Hagdu et al.'s work~\cite{hadgu_political_2013} studies the users leaning on a certain political side who start using political hashtags from the opposite side. It refers to these users as "hijackers". By combining statistical measurements of user volume with hashtags leaning towards political parties, this study aims at detecting a change point when a hijacking starts. Hijackers are often politically more active than the seed users of the hashtags. The authors in \cite{lansdall2016change} measure the change point and cause for the change in sentiment around the Brexit poll. This multivariate analysis uses the LARS algorithm on different regions in the UK and is temporally based on hourly windows. It analyzes five sentiments: anger, anxiety, sadness, negativity, and positivity using the Linguistic Inquiry and Word Count (LICW) lexicon. Positive sentiment decreases as other sentiments rise, often accompanied by external factors. In this case, the drop in the exchange rate has a significant co-relation with the mood swing.

Pita et al.~\cite{pita_linguistic_2016} show the linguistic characteristics of leaders and followers based on the Ecuadorian Twitter dataset of politicians and their followers. The authors evaluate twelve sentiments on the tweets in the LIWC software. Leaders have a different vocabulary set than followers, even though the followers' vocabulary is affected by the leader. The leader's tweet displays obvious certainty as compared to the tentativeness in the followers' tweets. Three categories (money, work, and achievement) are dominant in leaders linguistic features and are often found in Ecuadorian politicians. 

A study by Jain et al.~\cite{jain_politically:_2015} proposes a system to find politically similar friends for Twitter users based on their interactions on Twitter. It develops a recommendation algorithm that uses relatedness measures to construct a graph between different users. It uses mention, and re-tweet network along with content similarity in hashtags and content in a tweet.  The WalkTrap community detection algorithm is used on this graph to find communities of users with similar political interests.  
Meanwhile, Larsson and Moe~\cite{larsson2012studying} provide one of the earliest works on Twitter regarding user analysis. Their work is based on data collected for the Swedish election in 2010. The study shows that Twitter provides another platform for politicians and prominent position holders to disperse their information. The network participation is strongly affected by contextual factors, including elections.

Howard and Kollanyi~\cite{howard2016bots} study the use of political bots during a Brexit referendum in 2016. It shows that bots play an essential role in cascading misinformation on the Internet and are used strategically in conversations. While almost one-third of the content is generated by about one percent of the sampled accounts, such bots use different levels of automation for different contexts. In the Brexit setting, most of the bots retweeted the content with their own specific hashtags. 
Barberá and Rivero's  study~\cite{barbera2015understanding} shows that Twitter data has strong associations with the demographics settings. It analyzes user behavior in the Twittersphere of two different countries: the US and Spain, and its data spans the 70 days before the 2011 Spain's Legislative elections and the 2012 US General elections. It shows that for both demographics, male users are politically more active, and the opinions are highly polarized with an urban bias. 

Hosch-Dayican et al.~\cite{hosch2016online} analyze users' persuasive behaviors on Twitter during the Dutch election in 2012. It shows that such behavior is obvious among common users and political leaders. However, the sentiments of the conversations are different for each set of users. Politicians are more persuasive with their messages, while common users mostly use negative opinions to show their dislike.
Another study by Mustafaraj et al.~\cite{mustafaraj2009use} focuses on YouTube videos which deliever partisan messages. It also highlights the technological bias introduced by search engines on the ranking of the videos. Videos in the top 20 search results contained highly partisan messages for a period of six months before the US 2008 elections. Marozzo and Bessi~\cite{marozzo2018analyzing} propose a methodology to identify user polarization before elections. The study shows that news websites play a role in the polarization during the elections. However, opinionated users do not change their views, while the politically neutral users may sway to either side. 
Another method to look at influence is by measuring the power of politicians based on political debates. Prab-
hakaran et al.~\cite{prabhakaran_power_2013} analyze the Republican primary presidential debates for the 2012 elections. The authors use statistical analysis on features including the number of words used, questions asked, turns given to the political candidate, number of times a candidate is interrupted, number of times a candidate interrupts other candidates, the position and patterns of names and titles used in a debate, and the length of the debate. The authors also look at the lexical diversity and topics used in debates using LICW and LDA algorithms. The results show the positive co-relation among linguistic and topical usage and the candidate's position. Such Linguistic markers are also associated with trolling behaviors. Addawood et al.~\cite{addawood2019linguistic} focus on trolls on Twitter. The authors use LIWC to find the linguistic characteristics of trolls during the US Election and show that a higher number of hashtags, tweets, and retweets, along with a shorter length of messages with very few nouns are often associated with trolls profiles. This study identifies a large number of linguistic markers to train classifiers and predict troll users. 

In this section, we cover the work which focuses on the users and community modeling. The work focuses on identifying communities based on polarization, echo chamber, and homophily. In terms of individual users, the work focuses on finding political affiliation, gatekeepers, and influencer users. The most common methods for identifying the communities is to look at the information cascades, more particularly how users curate and share information across their networks such as retweet networks. Also, sentiment analysis is widely used to find the polarization in the content and networks. Features derived from statistical analysis on the volume of tweets and posts, TF and TF-IDF methods on the textual content, and sentiment analysis have been used both separately and as input to machine learning models for the classification tasks.

\begin{table}

\begin{center}
\caption{Work related to Influencer Users}

\begin{tabular}
{p{0.30\linewidth}p{0.15\linewidth}p{0.15\linewidth}p{0.28\linewidth}}
\toprule
\multicolumn{1}{c}{\textbf{Paper}} & \multicolumn{2}{c}{\textbf{Feature\footnote{`*' shows the feature used in method. No `*' sign in any of the columns shows that work is a qualitative study based on survey or interviews} }}  &\multicolumn{1}{c}{\textbf{Method}} \\
\toprule

  & \textbf{Linguistics} & \textbf{ Network} & \\
\cmidrule{1-4}

Rangenath et al 2016 \cite{ranganath2016understanding} &	*&	*&	LDA, Tensor \\
\hline

Jurgens et al. 2011 \cite{jurgens2011small} &	&	*&	Centrality Entropy \\
\hline

Hemsley et al. 2017 \cite{hemsley2017call} &	*&	*&	Classfication (SVM) \\
\hline

Hagdu et al. 2013 \cite{hadgu_political_2013} &	*&	&	Volume Analysis \\
\hline

Lansdall-Welfare et al. 2016 \cite{lansdall2016change} &	*&	&	LARS Algorithm, sentiement analaysis, LICW \\
\hline

Pita et al. 2016 \cite{pita_linguistic_2016} &	*&	*&	LIWC, Lingusitic Features \\
\hline

Jain et al. 2016 \cite{jain_politically:_2015}  &	*&	*&	Recommendation algorithm based on mention and retweet network \\
\hline

Larsson and Moe 2012 \cite{larsson2012studying}&	*&	*&	exploartory \\
\hline

Howard and Kollanyi 2016 \cite{howard2016bots} &	*&	&	Exploartory \\
\hline

Barbera and Rivero 2015 \cite{barbera2015understanding} &	*&	&	Exploartory, sentiment\\
\hline

Hosch Dayican et al. 2016 \cite{hosch2016online} &	*&	&	Sentimen Analysis \\
\hline

Mustafaraj et al. 2009 \cite{mustafaraj2009use} &	&	&	exploartory, videos messages\\
\hline

Maarzo and Bessi 2018  \cite{marozzo2018analyzing} &	*&	&	Linguistic and Polarization Analysis \\
\hline

Prabhakaran et al. 2013 \cite{prabhakaran_power_2013}&	*&	&	Statistical analysis, LICW, LDA\\
\hline

Addawood et al. 2019 \cite{addawood2019linguistic} &	*&	&	LICW\\

\bottomrule

\end{tabular}
\label{t_influencer_users}
\end{center}
\end{table}

\subsection{Information Flow}
The quality of information and its propagation affect the analysis and results of the computational methods. Online content is highly prone to noise and fake information due to the lack of specific moderation. In this section, we discuss the work that focuses on different aspects of political misinformation, from biased to fake news. In particular, as social media and most of the online communities allow the sharing of data from third party websites it is important to consider the information quality, and its dissemination patterns. 


Political news are useful mediums for conveying messages; a study by Brewer et al.~\cite{brewer2013impact} demonstrates that exposure to the news is positively co-related to a  higher critical opinion. Such exposure to information can easily affect people's opinions. 
Research in the news itself is a broad area and can range from detecting fake and hoax news, identifying propaganda, and finding satire in the news particularly in political entertainment media sources. Most common techniques involve using linguistic features from the text, creating and analyzing the network of words and hyperlinks. However, the credibility of the news is a huge problem when there is so much data. Several research works have focused on the authenticity and detection of fake news along with finding political bias in the media. 
   
A study by Lumezanu et al.~\cite{lumezanu2012bias} uses two datasets related repectively to the Nava Senate Race in 2010 and the debt-ceiling crisis in 2011\footnote{\url{https://en.wikipedia.org/wiki/United\_States\_debt-ceiling\_crisis\_of\_2011}}  debates to analyze  propaganda dissemination on Twitter. The authors identify certain characteristics of propagandists such as the sending of a high volume of tweets, colluding with other, quickly retweeting, and retweeting without adding any of their own content. Kim et al.~\cite{kim2018leveraging} introduce a framework called "CURB" based on crowdsourced flagging to detect fake news and prevent  misinformation. However, the system considers every user with equal probability for being good or bad. Ciampaglia et al.~\cite{ciampaglia2015computational} display the usage of Wikipedia for automated fact-checking. The authors use a network-based approach by treating the Wikipedia Knowledge Graph as a distant graph. This study shows that distance measures on such a graph can be used for fact-checking. 
Gyöngyi et al.~\cite{gyongyi2004combating} propose a method to filter the  spam web pages with a semi-automatic technique based on the page rank method. Shu et al.~\cite{shu2018studying} discuss three dimensions of news flows in networks; content, social, and temporal. The study features user network properties such as echo chamber and filter bubble, diffusion, friendship, and credibility network along with heterogeneous networks such as knowledge graphs for identifying and reducing the diffusion of fake news. 
Wu and Liu~\cite{wu2018tracing} classify the fake news and rumors in social media and proposes TraceMiner, a machine learning-based method  that uses the propagation of messages in the network. It adds the user embedding to the network and achieves up to 91 percent precision. Dong et al.~\cite{dong2015knowledge} exploit the reliability of sources to propose a knowledge-based trust model, which depends on the amount of correct factual information provided by the data source. The model also measures the accuracy of the extraction and data itself. \yd{Besides, Morgan et al.~\cite{morgan_is_2013} examine the effects of the perceived ideology of news outlets on the consumption and sharing by studying news using Twitter users. The authors find that as users share news from more diverse news outlets, they tend to quickly incorporate news content with opposing viewpoints.}
Finally, another study~\cite{dong2014data} uses data fusion techniques to solve the knowledge fusion problem for fact-checking subject-predicate-object triples.   

Bias can be introduced into the system through different methods. Kulshrestha
et al.~\cite{kulshrestha2017quantifying} present a framework to quantify the sources of political bias in Twitter from three sources: (1) \textit{Input}: bias in the data that is fed into the ranking system, (2) \textit{Output}: effective bias presented to a user, and (3) \textit{Ranking Bias}:  add-ons to the input data bias  introduced by the ranking system. The authors also show that the input data and the ranking system play an important role in the search results. Robertson et al.~\cite{robertson_auditing_2018} conduct an audit on Search Engine Result Pages (SERP) of political queries on the Google search engine. They show that search queries impact the information path and have an effect on the result aggregation. It also affects the personalization of the result. 

Gentzkow and Shapiro~\cite{gentzkow2010drives} study the bias in a leading US newspaper by finding a positive co-relation between the keywords used by politicians in their debates and the words used by newspapers in the news reports for those debates. However, the authors did not consider the newspaper' generated content. Another  work~\cite{efron_liberal_2004} based on a study by Turney and Littman~\cite{turney_unsupervised_2002} uses the document and hyperlink co-citations to find the political bias in the web documents.  
Jackson et al.~\cite{jackson2017identifying} use a lexicon based approach on Twitter and Facebook messages to find the political topics discussed in users' chats.  Finally, a last study by Qi et al.~\cite{qi2017social} uses machine learning algorithms on tweets of political leaders to find policy topics set by Political Agenda Projects with 78 percent accuracy. 

Studying the behavior of Facebook users regarding political news sharing and consumption~\cite{an2013fragmented} shows that partisan selectiveness exists for social network users. It can lead to the formation of echo chambers in thesocial media and increase the bias in feeds for the users in a network. 

Additional exposure to information helps to enhance the social experience. Users can be passively receptive to the information, or they can participate in the process. In the case of Twitter, some users just observe the tweets from other people, while some other users tweet their opinions. Being receptive or active does not matter in users feeling towards politicians nor does it affect what has already been learned by users regarding politicians~\cite{maruyama2014hybrid} . However, active tweeting behavior influences the vote choice compared to receptive users or non-Twitter users. 

Alfina et al.~\cite{alfina_utilizing_2017} use  hashtags from tweets to train  binary classifiers and predict the political sentiments in tweets. The authors show that the use of hashtags increases the accuracy of the classifiers. The method only uses the hashtags that are political in nature and name those as sentiHT. The dataset resulting from this method provides better accuracy than manually labeled tweets. When combining all of the features, unigram (build on unique words in tweets), sentiHT, and sentiLex (based on lexicon built on the Indonesian language),  Random Forest and Logistic Regression give the best results with an accuracy of 97 percent. Finding sarcasm and irony in tweets is more challenging than finding general political sentiments. Similarly, Charalampakis et al.~\cite{charalampakis_detecting_2015} use a Twitter dataset from Greece to predict the irony in political tweets. The study uses five features from the tweet text: Verbal, Rarity, Meaning, Lexical, and Emoticons to train classifiers. The different classification algorithms achieve a significant precision in results, with the Tree-Based algorithm leading with an 82.4 percent precision score.

While the above works focus on the political bias of users, Loyola et al.~\cite{loyola_using_2016} predict the tendency of the legislative bill in favor of either corporations or general good. It uses TF-IDF, LDA, and Word Embedding methods to build the anti and pro corpus for the bills on the legislative assembly data and web data during two consecutive legislature periods from 2006 to 2014 in Chile. 

Active users of the system can also introduce  bias by adding irrelevant data to post to distract the attention from the main topic to another topic. Wang et al.~\cite{wang_diversionary_2012} develop a framework to detect the diversionary comments on political blogs. The method is based on textual features and involves reference resolutions, Wikipedia's first paragraph for more data points to the topics, and LDA. It achieves a mean average precision (MAP) of 89.5 percent. Another study by Krestel et al.~\cite{krestel_treehugger_2012} uses TF-IDF on a corpus of four major German newspapers to detect the bias in  political news. It shows that newspapers tends to use certain vocabulary for certain political groups. 

A paper by Garrett and Weeks~\cite{garrett_promise_2013} shows that merely providing correct and factual information is not sufficient enough to remove the perception of users, particularly the perception on contentious topics. It shows that a medium interaction design and strategy to deliver information plays an equally important role. 

This section covered the work mainly on quality of information flow such as misinformation, political bias, and fact checking. For political bias the most common method used to find sentiment, is to use lexicon base approaches for political topics. Methods to find fake news and fact checking focus on the publishing patterns of users and building the knowledge graph with heterogeneous data sources. 

\begin{table}
\caption{Studies Focused on Information Flow }
\begin{center}
\begin{tabular}
{p{0.30\linewidth}p{0.15\linewidth}p{0.15\linewidth}p{0.28\linewidth}}
\toprule
\multicolumn{1}{c}{\textbf{Paper}} & \multicolumn{2}{c}{\textbf{Feature\footnote{`*' shows the feature used in method. No `*' sign in any of the columns shows that work is a qualitative study based on survey or interviews} }}  &\multicolumn{1}{c}{\textbf{Method}} \\
\toprule

  & \textbf{Linguistics} & \textbf{ Network} & \\
\cmidrule{1-4}

Lumezanu et al 2012 \cite{lumezanu2012bias}	& *& *& Tweet Volumes, Term Frequency \\
\hline
Kim et al. 2018 \cite{kim2018leveraging}	& *& &crowd sources system \\
\hline
Ciampagla et al. 2015 \cite{ciampaglia2015computational}	& &* &Graph based, distant graph on Wikipedia Knowledge Graph\\
\hline
Gyongyi et al. 2004 \cite{gyongyi2004combating}& & *&	page rank method\\
\hline
Shu et al. 2018	\cite{shu2018studying} &* & *&Network Properties\\
\hline
Wu and Liu 2018 \cite{wu2018tracing}	& &* &Machine Learning - LSTM - RNN together with network structure information\\
\hline
\yd{Morgan et al. 2013 \cite{morgan_is_2013}}	& & & exploratory\\
\hline
Dong et al. 2014 \cite{dong2014data}	&* & &Data Fusion Technique\\
\hline
Dong et al. 2015 \cite{dong2015knowledge}	& &* &Knowledge based trust\\
\hline
Kulshreshta et al. 2017 \cite{kulshrestha2017quantifying}	&* & &quantification framework  for social media search engines\\
\hline
Robertson et al, 2018 \cite{robertson_auditing_2018}	&* & &Search Path auditing\\
\hline
Efron 2004	\cite{efron_liberal_2004}& *& *&co-citation analysis\\
\hline
Jackson et al. 2017	 \cite{jackson2017identifying}&* & &Lexicon Based Aproach\\
\hline
Qi et al. 2017 \cite{qi2017social}	& *& &Machine Learning\\
\hline
An et al. 2013\cite{an2013fragmented} 	& &* &\\
\hline
Maruyama et al. 2014 \cite{maruyama2014hybrid} & & &	\\
\hline
Alfina et al. 2017	\cite{alfina_utilizing_2017}& *& &binary classifiers RF and LR\\
\hline
Charalampakies et al. 2015 \cite{charalampakis_detecting_2015}&* & &	Classification on tweet features\\
\hline
loyola et al 2016 \cite{loyola_using_2016}	& *& &TF-IDF, LDA, Word Embedding\\
\hline
wang et al. 2012	\cite{wang_diversionary_2012}& *& &reference resolution on textual reference\\
\hline
krestel et al. 2012	\cite{krestel_treehugger_2012}& *& &TF-IDF \\
\hline
Gerrett and weeks 2013 \cite{garrett_promise_2013} & & &	exploratory\\

\bottomrule

\end{tabular}
\label{t_information_flow}
\end{center}
\end{table}

\subsection{Political Discourse}
This section consists of the work related to the political discourse of  users and how it affects user behaviors in terms of political activism. Prominent approaches include studying the network structure, community effect, and users interactions with the rest of the network.

Balasubramanyan et al.~\cite{balasubramanyan2011pushes} use machine learning algorithms to predict the emotional response of the community to blog posts based on the political topics in the blog text. This study shows that certain political topics result in higher collective sentiments from the community. These emotions vary with the community. Pointwise Mutual Information (PMI)~\cite{turney2002thumbs} gives better accuracy than SenitwordNet. The authors show that the community context plays a significant role in the performance of machine learning algorithms on blog posts of different political spectrums. 
A study by Mejova and Srinivasan~\cite{mejova2012political} suggests that the type of social media platform is a significant factor in deciding the political sentiments and consensus results among the users. It compares YouTube and Twitter for political sentiments on similar topics and queries. Results show that there is no similarity between sentiments and agreements. It also finds that YouTube has a higher percentage of off-topic discussions than Twitter.  Another study~\cite{mejova2013gop} shows that Twitter does not respond to the national polling results. The authors train and test a machine learning algorithm for political sentiment analysis. It uses a two-step approach to find the pro, against and neutral sentiments in the first step. After that, it uses all the scores to predict the political sentiments. The method uses tuning on threshold adjustments for the range of these sentiments, with increased accuracy. The authors also show that off-the-shelf machine learning algorithms do not give significant results. 

Mohamed et al.~\cite{mohamed2017private} show through a survey of social media users that online activities affect the political candidates. This study analyzes how political candidates' activities are seen by users compared to other types of profiles. Users are more critical in the case of a political candidate than for a job seeker's profile. However, this tolerance level is not consistent in different demographic groups. Another study by Hoffmann and Lutz~\cite{hoffmann2017spiral} show that Facebook users' social networks affect their political engagement. Users under 30 years old try to engage less in political debate in case of heterogeneous political opinions in the network. However, they also suggest some deviation based on the role of social network personalized feed algorithms to present users with more related information and showing less heterogeneity in the network.

Zhu et al.~\cite{zhu_multidimensional_2009} show the importance of feature selection in political opinion mining and how it is different from typical opinion mining tasks. They use Principal Component Analysis (PCA) and Regression algorithms for analysis. Marchetti-Bowick and Chambers~\cite{marchetti2012learning} use a distance learning approach to identify  political topics with a 0.91 F1 score. The authors also find a high correlation between Twitter data and Gallup rating surveys for political forecasting. 
Another interesting area in political discourse is contrastive opinion mining (COM). Fang et al.~\cite{fang2012mining} introduce a novel COM algorithm to compare and contrast opinion distributions within both US senate debates and newspaper articles from China, India, and the USA. A study by Stier~\cite{stier2016partisan} shows that the framing of tweets is observable, between politicians with different partisans, based on topics and linguistics .
Semantic shifts are also observable in contrastive opinions
on topics. Azarbonyad et al.~\cite{azarbonyad_words_2017} proposes a method based on word embeddings to measure the semantic similarity in political viewpoints. It shows that semantic shift is not only a temporal phenomenon but also happens with different point of views. Results show a 0.74 F1 score. Another study~\cite{trabelsi2018unsupervised} introduces an unsupervised learning model to discover the interaction discourse and viewpoints of authors on posts from online discussion forums. It considers the difference of opinions as causals of interaction among authors. A last work by Trabelsi and Zaine~\cite{trabelsi2019phaitv} uses an approach based on topic interaction viewpoint to find and summarize the arguments used in discussions. 

Political disaffection is the phenomenon which shows people's lack of confidence and dissatisfaction.  Monti et al.~\cite{monti2013modelling} analyze political disaffection for the first time on Italian Twitter users. The study applies machine learning models to  tweets that have political biases, negative sentiments and are not directed towards certain politicians or political parties. The study shows that political disaffection on Twitter is a temporal reflection of the results of a survey conducted  in the same period. It also suggests that Twitter data can be used as a valid measurement of such phenomenon. 

In 2015, Twitter allowed users to add content when they retweeted someone else's post~\cite{pocket-lint_how_2015}. This kind of interaction is called \textit{'Qoute Retweet'}. This new interaction method affects the political discourse of the users and leads information diffusion to occur earlier~\cite{garimella_quote_2016}. Early adopter of this kind of usage are comparatively active users and sentiment of the discussions discourse stays positive.
Kannangara~\cite{kannangara_mining_2018} proposes a probabilistic classifier for socio-political opinion polarity using the three-dimensional scheme Joint-Entity-Sentiment-Topic(JEST). JEST finds user ideology towards political topics and entities. It uses LDA based algorithms to find entities, topics, and sentiments. This study also proposes a context based sarcasm detection model built upon the first two models. 

Topical analysis of political content can be used to find the latent features behind the use of such topics. Greene and Cross~\cite{greene_unveiling_2015} apply a two-layer matrix factorization approach for dynamic topic modeling on English language speeches from European Parliament members. Speeches are divided into coherent time frames to get dynamic and temporal study data. The study considers these speeches to be a single topic. To find more than one topic in a particular time window, the authors use topic co-relation. Temporal usage of topics compared with different case studies shows that the choice of the topic used by parliament member relies heavily on thier partisan orientation, and the position of the member in the parliament. Ardehaly and Culotta~\cite{ardehaly_mining_2017} mine the demographics and opinions from Twitter data. They propose an algorithm, weighted label regularization, that scales the earlier model of learning using label proportions (LLP). This work also uses conditional probabilities between users’ latent attributes. The authors suggest their method can supplement polling results. 

Verma and Ramamurthy~\cite{verma_analysis_2016} develop a framework with a series of algorithms to find suggestion, opinion, and sarcasm from political blogs using textual features. The authors use natural language processing libraries to find the part of speech (POS) from the comments to find if there are verbs or suggestive words in the comments. They use discourse connectors to find the connections between two sentences. Once the connection is established, they accordingly calculate the polarity  and extract the opinion. For sarcasm, they compare the sentiments of all sentences in the pair. Reported accuracy for this framework is 86 percent.

Apart from looking for opinions and suggestion, social media can help in Political Crisis Detection (PCD). Keneshloo et al.~\cite{keneshloo_detecting_2014} use the Global Dataset of Events, Language, and Tone (GDELT) dataset that is based on the news article for events across the world to predict local political crisis. It uses LogitBoost and LibSVM libraries for classificatying whether the local graph is either PDC or not. For prediction, it proposed three models: the first model is based on the event count in a monthly time period and the event properties; the second model considers the network properties; the third model is an hybrid between the first two models. For prediction, the latter model achieves a precision of up to 0.95. 
Chan and Fu~\cite{chan_predicting_2015} manage to predict cyber balkanization based on Facebook data collected during the Hong Kong political crisis in 2014. It shows that, by using network sharing ties, the younger generation, which is more active on Facebook than the older generation in HK, is more likely to be politically polarized and cyber-balkanized.

Primario et al.~\cite{primario_measuring_2017} show that political conversation on Twitter is pruned to polarization and the U-turn effect: the degree of polarization is high at the start and at end of the conversation while it lowers in the middle. It suggests such behavior is due to a higher degree of engagement with other users towards the middle of the conversation. The polarization detection framework consists of the adjacency matrix of re-tweet networks, elite user detection, and the measure of polarization using the method proposed by Morales et al.~\cite{morales2015measuring} . It defines elite users who have at least 100000 followers, are related to politics, and participate significantly. To predict political polarization, elite users are more important than common users or listeners. Another study~\cite{an2019political} uses Reddit to analyze the interaction and linguistic characteristics of users in the homogeneous and cross-cutting communities. The analysis is based on the interaction between proponents of politicians. They consider linguistic similarities such as style, vocabulary, and semantics, using already proposed methods. The results show that heterogeneous communities have cross-cutting political discussions among supporters of different politicians. The authors show that linguistic style in general is not democratic, but users adapt their style according to posts. The study also displays the lack an echo chamber effect on users from the observed communities.

Makki et al.~\cite{makki_active_2015} propose an active labeling framework to relate tweets to political debates. The framework is modeled and trained on Canadian political debates and Twitter users. Along with using unsupervised learning, it focuses on active labeling where the model focuses on the tweets with near-similar scores more than one tweets that are rated equally for some topic, politically related hashtags to the debates, and the tweet with unusual discourse in terms of replies. It shows that this active labeling method has better accuracy and a combination of all these selection criteria can achieve up to 90 percent accuracy.  Another study by Joseph et al.~\cite{joseph2019polarized} analyzes the response of partisan users to Trump's tweets. It found that there are agreements between supporters of each party aside from the absolute difference. Le et al.~\cite{le_revisiting_2017} study the American bipartisan Twitter users, around the 2016 Presidential elections, in three dimensions: Personality traits of candidate, Party, and Policy. The authors conduct an exploratory study to reinforce that these traits are still important for deciding upon voting behavior, initially studied through surveys in 1960 and then in 2008~\cite{noauthor_american_nodate,lewis-beck_american_2008} respectively. The authors used Adjective Check List (ACL) along with personality templates to identify personality traits of candidates and the  SENTIWORDNET lexicon to measure the sentiments in tweets. Statistical analysis with very low p-values shows the significance of results for inter and cross-partisan behavior.

\yd{
Hemphill et al.~\cite{hemphill_whats_2013} investigate how officials utilize Twitter to advertise their political opinions based on data from 380 members of the US Congress' Twitter activities during the winter of 2012.
State and Adamic~\cite{state_diffusion_2015} examine complex diffusion characteristics congruent with threshold models. The authors find that the adoption probability depends on the number of friends and the susceptibility of the individual. 
In addition, Semaan et al.~\cite{semaan_navigating_2015} interview 27 residents in the state of Hawaii who use at least one social media tool to participate in the online public sphere. The authors examine why and how people use multiple social media for political involvement.
Lastly, Savage and Monroy-Hernandez~\cite{savage_participatory_2015} conduct a statistical analysis of a Facebook page, so-called "VXM" ("Valor Por Michoacan" Spanish for "Courage for Michoacan"), and describe VXM's online mobilization strategies.
}

While most of the work focuses on English language content, a few works study politics in different languages. Elghazaly et al.~\cite{elghazaly2016political} show that the Naive Bayes classifier on the Arabic language corpus from Egyptian Twitter users gives the best accuracy and precision rate of 92.5 percent to find the political sentiments. It also suggests that tweets may represent the immediate thinking of users to certain topics or phenomenon to spread their ideas. Another study~\cite{boireau2014determining} studies the Twitter timelines for Belgian politicians and suggests that their profiles can allow them to deduce an accurate political ideology.  Jürgens et al.~\cite{jurgens2011small} analyze the political conversations happening during the German general election of 2009. They observe that, by using entropy measures, certain positions (Gatekeepers) in the network are more important than others in regard to information diffusion for the rest of the network. 
Sandoval-Almazan and Valle-Cruz~\cite{sandoval-almazan_facebook_2018} studies the emoticons on Facebook profiles for the Mexican 2017 elections. Through an empirical study, they find that positive Facebook emoticons do not have a positive relationship with the winning party. However, the sample of their study is only 4128 posts, so data is limited with demographics characteristics being the main limitations of the dataset. 
Calderon et al.~\cite{calderon_mixed-initiative_2015}  show that Twitter does not provide any evidence of a correlation between negative sentiments on Twitter and street protests in Brazil after the Football world cup. However, Twitter data does support the grievance theory where one feels deprived based on a comparison with someone. 
Finally, a study by Yonus et al.~\cite{younus_what_2011} uses Twitter data to find the political subjectivity in tweets during the political crisis in Tunisia. It shows that social features like follower/following ratio, user mentions, lists, and textual features similarity in the tweets can be used to predict the subjectivity in the tweet. A system trained on these features using a Naive Bayes classifier achieves 83 percent accuracy.  

We discussed work related to political discourse in this section. Work in this area ranges from observational and qualitative surveys to statistical analysis of interaction patterns and data. Studies mentioned in this section focus on opinion forming, interaction patterns among different political entities and users, and analysis of political content. Work in this section relies heavily on linguistic features and used classification algorithms to find topics and opinions of users. 

\afterpage{
\begin{center}
\begin{longtable}[h]{p{0.30\linewidth}p{0.15\linewidth}p{0.15\linewidth}p{0.28\linewidth}}
\caption{Work Focused on Political Discourse} \label{t_political_discourse} \\

\toprule
\multicolumn{1}{c}{\textbf{Paper}} & \multicolumn{2}{c}{\textbf{Feature\footnote{`*' shows the feature used in method. No `*' sign in any of the columns shows that work is a qualitative study based on survey or interviews} }}  &\multicolumn{1}{c}{\textbf{Method}} \\

& \textbf{Linguistics} & \textbf{ Network} & \\
\cmidrule{1-4}


\endhead

\hline \multicolumn{4}{|r|}{{Continued on next page}} \\ \hline
\endfoot

\endlastfoot

Balasubramanyan et al. 2011 \cite{balasubramanyan2011pushes} &*&*&Pointwise Mutual Information \\
\hline
Mejova and Srinivasan. 2012\cite{mejova2012political} &*&& Sentiment and Statistical Analysis\\
\hline
 Mejova et al. 2013\cite{mejova2013gop}&*&&SVM\\
 \hline
Marchetti-Bowick and Chambers. 2012\cite{marchetti2012learning} &*&&Distance Learning Approach\\
\hline
Mohamed et al. 2107\cite{mohamed2017private}&&&Survey\\
\hline
Hoffmann and Lutz. 2017\cite{hoffmann2017spiral}&&&Survey\\
\hline
Zhu and Mitra. 2009\cite{zhu_multidimensional_2009}&*&&Regression Analysis, PCA\\
\hline
 Fang et al. 2012\cite{fang2012mining}&*&&COM Algorthims\\
\hline
 Stier. 2016\cite{stier2016partisan}&*&&Linguistic Analysis\\
\hline
 Azarbonyad et al. 2017\cite{azarbonyad_words_2017}&*&&Word Embedding\\
\hline

Trabelsi et al.\cite{trabelsi2018unsupervised}&*&*&Unsupervised Learning on Author Topic Interactions Model\\
\hline
Trabelsi and Zaiane. 2019\cite{trabelsi2019phaitv} &*&*&Topics Interactions Model\\
\hline
Monti et al. 2013\cite{monti2013modelling}&*&& Political Dissaffection Measurement using Machine Learning\\
\hline
Garimella et al. 2016\cite{garimella_quote_2016}&*&*&Conent Analysis\\
\hline
Kannangara. 2018 \cite{kannangara_mining_2018}&*&&Joint-Entity-Sentiment-Topic Model\\
\hline
Greene and Cross. 2015\cite{greene_unveiling_2015}&*&&Two-layered Matrix Factorization\\
\hline
 Ardehaly and Culotta. 2017\cite{ardehaly_mining_2017} &*&&weighted label regularization\\
\hline
 Verma and Ramamurthy. 2016\cite{verma_analysis_2016} &*&&Natural Language Processing, Sentiment Analysis\\
\hline
 Keneshloo et al. 2014\cite{keneshloo_detecting_2014} &&*&LogitBoost  LibSVM \\
\hline
 Chan and Fu. 2015\cite{chan_predicting_2015}&&*&Network Aanalysis\\
\hline
 \yd{Hemphill et al. 2013\cite{hemphill_whats_2013}} & * & * & Empirical Study \\
\hline
 \yd{State and Adamic 2015\cite{state_diffusion_2015}} &  & * & Empirical Study, Diffusion Model\\
\hline
 \yd{Semaan et al. 2015\cite{semaan_navigating_2015}} &  & & Interview\\
\hline
 \yd{Savage and Monroy-Hernandez 2015\cite{savage_participatory_2015}} & * & & Empirical Study\\
\hline
 Elghazaly et al. 2016\cite{elghazaly2016political}&*&&Naive Bayes classifier\\
\hline
 Boireau. 2014\cite{boireau2014determining}&*&&\\
\hline
 Jurgens et al. 2011\cite{jurgens2011small}&*&*&Entropy Measures\\
\hline
 Sandoval-Almazan and Valle-Cruz 2018\cite{sandoval-almazan_facebook_2018}&*&&Empirical Study\\
\hline
 Calderon et al. 2015\cite{calderon_mixed-initiative_2015}&*&&Sentiment Analaysis, Deprivement Theory\\
\hline
 Younus et al. 2011\cite{younus_what_2011}&*&&Subjectivity Prediction with Naive Bayes classifier\\
\hline
 Primario et al. 2017\cite{primario_measuring_2017}&&*&polarization detection framework\\
\hline
 An et al. 2019\cite{an2019political}&*&*& Linguistic and Interaction Aanalysis\\
\hline
 Makki et al. 2015\cite{makki_active_2015} &*&&Active Label Framework\\
\hline
Joseph et al. 2019\cite{joseph2019polarized}&*&*& Exploratory\\
\hline
Le et al. 2017~\cite{le_revisiting_2017}&*&*&Sentiment and Statistical Analysis\\
\bottomrule
\end{longtable}
\end{center}
}

\subsection{Election Campaigns} 

In this section, we highlight studies that focus on the role of media in political events such as election campaigns and result prediction. In every democratic country, elections are the prime of democracy, and everybody is curious to know the outcome of polls. Political parties are interested in gauging their success rate and focus on their own political campaigns. Many other organizations, including mass media, are also interested in such predictions. Traditionally, user surveys are conducted in different regions to get people's opinions. Mass tendencies towards certain political parties are measured based on these opinions. The party with the higher tendency is predicted to be the winner.  However, the scale of such surveys is not large, and surveys are dependent on a large workforce, requiring a lot of time and cost. Recently the online media, mainly social media platforms, have given an alternative to traditional surveys and polls, and it has become very convenient to get a broader range of opinions from different demographics.

However, such platforms have limitations compared to polls and surveys that have specially designed questions. Social media data is more about users' personal opinions and self-information with various explicit and latent attributes and data. Computational methods need to be applied to these media to extract the information needed and gauge users' political tendencies.



\subsubsection{Political Campaigns}
Social media provide an excellent platform for engaging users in political discussions, rapidly spreading information and have been used by politicians and organizations during the political process. In some cases, social media has been used to organize the protests that have led to a change in government. In Section~\ref{subsect: homophily}, we have already mentioned the \textit{\enquote{tiny act of donation}} make it easier for users to participate.

Wulf et al.~\cite{wulf2013fighting} show that social media, among other technological structures, can be used for political engagement by discussing how Facebook and YouTube spur the protests in a Palestinian village. A study by Felt~\cite{felt2016mobilizing} shows how Twitter was used to start the inquiry for missing and murdered women in Canada. It was also one of the reasons for people voting opinions. Equality of political voice is very important in a democracy. However, political interest groups such as lobby groups try to have a louder political voice. Hong and Nedler~\cite{hong2015social} study the effect of the Twitter network on political voice. Even though social media brings huge diversity, it strengthens the political voice relative to the size of the network of followers of some organizations. This study also finds that a larger organization with a bigger number of Twitter followers has a louder political voice, although it does not answer why such a higher concentration occurs.

Saleiro et al.~\cite{saleiro_sentiment_2016} predict  political polls by building prediction models based on sentiments of the tweet related to the Portuguese bailout (2011-2014). The model uses aggregate sentiment functions with univariate and recursive feature selection methods. This study shows that the \textit{bermingham} feature is the most significant. The evaluation results in a Mean Absolute Error (MAE) of 0.63\%.

Suarez et al.~\cite{hernandez-suarez_predicting_2017} show the co-relation between Twitter sentiments and political preferences in a certain period. The findings however are not generalizable for overall prediction. The authors use a classifier based on the Naive Bayes classifier on labeled tweets during the US Election  in 2016. The probabilistic model considers sentiment, features, numbers of features, and the frequency of each feature in training.

A study by Raja et al.~\cite{raja_detecting_2016} shows how information flows in a Twitter network according to the political situation in a country. The authors compare data from politically unstable and stable countries to analyze the flow and show that re-tweet and closely connected loops are more common in a Twitter network in politically unstable countries. They also use information cascades, Naive Bayes, and SVM for classification and report up to a 100 percent true positive rate.

Smailovic et al.~\cite{smailovic2015monitoring} use SVM classification for binary sentiment analysis and three-class sentiment analysis on two sets of tweets collected around the Bulgarian parliamentary elections in 2013. Both sets of tweets include general and political tweets. Result shows that the political sentiment changes as the election comes closer, and contextual factors such as political scandals affect the sentiments as the election comes closer. The proposed model is evaluated at a 10-fold validation and is accurate up to 90 percent on different n-grams evaluation. 
Another work by Sukul et al.~\cite{sukul2017online} studies the micro-targeting of political ads to online viewers. The authors develop a bot system to analyze the on-screen advertisements on YouTube videos. They show that adds appearance does not differ significantly for human and bot users. However, the ratio of political advertisements is higher in battleground states; the states that have higher chances of winning for both the parties \footnote{\url{https://en.wikipedia.org/wiki/Swing\_state}}, as compared to that in non-battleground states. The ratio of appearance also depends on the political events.

\begin{table}
\caption{Studies Focused on Political Campaigns}
\begin{center}
\begin{tabular}
{p{0.30\linewidth}p{0.15\linewidth}p{0.15\linewidth}p{0.28\linewidth}}
\toprule
\multicolumn{1}{c}{\textbf{Paper}} & \multicolumn{2}{c}{\textbf{Feature\footnote{`*' shows the feature used in method. No `*' sign in any of the columns shows that work is a qualitative study based on survey or interviews} }}  &\multicolumn{1}{c}{\textbf{Method}} \\
\toprule

 & \textbf{Linguistics} & \textbf{ Network} & \\
\cmidrule{1-4}
Wulf et al. 2013\cite{wulf2013fighting} &&&Interview, Observational\\
\hline
Felt. 2016\cite{felt2016mobilizing}&*&&Qualitative\\
\hline
 Hong and Nadler. 2015\cite{hong2015social}&&*&Empirical Analysis\\
\hline
Saleiro et al. 2016\cite{saleiro_sentiment_2016} &*&&Sentiment Analysis\\
\hline
Suarez et al. 2017\cite{hernandez-suarez_predicting_2017} &*&&Naive Bayes classifier\\
\hline
Raja et al. 2016\cite{raja_detecting_2016} &&*&information cascades, Naive Bayes, and SVM\\
\hline
Smailovic et al. 2015\cite{smailovic2015monitoring}&*&&multiclass SVM\\
\hline
Sukul et al. 2017\cite{sukul2017online}&*\footnote{content related to ads on YouTube}&&Content Similarity Analysis\\

\bottomrule

\end{tabular}
\label{t_election_campaigns}
\end{center}
\end{table}

\subsubsection{Politics and Election Prediction}


Different social mediums and social networks offer different kinds of opportunities for users to interact and share their thoughts. In a study published in 2018, Siegel~\cite{siegel_life_2018} discusses the role of social networks in users' political understanding. The study finds, using statistical analysis on survey data, that Twitter users find themselves to have a better political understanding as that of Facebook users. Hoffmann et al.~\cite{hoffmann2017facebook} use Linear Regression to find that Facebook users can have a positive relation with online political participation
even through accidental engagements with political posts. Semaan et al.~\cite{semaan2014social} show in a qualitative study that once the users are provided with diverse options for deliberative discussions, users prefer to choose mediums where they could express their views more effectively. However, some users can also change their views during the course of such interactions. 
Agarwal et al.~\cite{agarwal2019tweeting} study the interaction characteristics of the UK members of parliament (MP) and other Twitter users based on tweets, replies, mentions, and followers/following connections. The analysis shows that politicians use selective strategies to reply to certain users from the UK. However, Twitter users have a short and bursty period of mentions for politicians that usually lasts for around three days. Then, the focus period moves to other politicians. The sentiment analysis performed by LIWC shows that tweets from users to politicians often contain strong and moralizing words depending on the context and past event, for instance, a scandal. However, there is a positive sentiment between MPs and citizens.   
Kulshrestha et al.~\cite{kulshrestha2017politically} studied the role of Twitter in the 2014 Indian election campaigns. The study shows through statistical measures on information diffusion and augmented contagion model on the Twitter network that the winning party has run a successful campaign using social media. 
A study by Li et al.~\cite{li2018working} shows how nonprofit organizations (NPO) use Twitter in politically critical situations for immigrants. The authors utilize the Empowerment Theory combined with statistical measures on users’ profile attributes and LIWC. They observe that such an organization uses three strategies: diffusing more information, calling for collective participation from the rest of the users, and communicating with outside actors including politicians, governments, and media.  The study highlights the design challenges to reduce the communication gap between outside actors and NPOs while keeping the surveillance effect to a minimum. Hassanali and Hatzivassiloglou~\cite{hassanali2010automatic} use SVM  to train a classifier on web blog data to predict the tags for political posts. The authors also perform name entity extractions from the post. Tagging political posts is different from general posts. Extraction of the named entity helps in this process. This work achieves a recall of 75 and 65 percent on Daily Kos\footnote{\url{https://www.dailykos.com}} and Red State Data \footnote{\url{https://www.redstate.com}}, respectively.

Measuring sentiments in the content has been one of the prominent technique to co-relate it with voting tendencies. Oliveira et al.~\cite{oliveira2017can} use the sentiment analysis on tweets to find the political preferences and compare it with the users’ survey results. The authors find that sentiment analysis results vary ranging from 1 to 8 percent of the traditional opinion polls conducted by organizations. The accuracy of traditional opinion poll lies at an average of 81.05 percent. Tumasjan et al.~\cite{tumasjan_predicting_2010} present one of the earliest works to study political results based on Twitter data. The work analyzes 12 dimensions of sentiments using LIWC to profile the political sentiments, and based on those analyses, it shows that such sentimental profiles can be real-time measures of the political situation and voting tendencies. A study by Kagan et al.~\cite{kagan_using_2015} used Twitter data to predict the election outcome in the Pakistani and Indian general election in 2013 and 2014, respectively. The proposed method uses the AVA sentiment analysis algorithm to measure the sentiments for the leading candidates contesting for the leadership. The result shows that the winning candidates from both countries had a higher positive sentiment than the competing candidates.  Dokoohaki et al.~\cite{dokoohaki_predicting_2015} use a network-based approach to study the links formed based on the conversation threads that discuss politicians. The study uses stochastic link prediction algorithms to find and co-relate the political accounts having stronger links with the outcome of the elections. 
Tsakalidis et al.~\cite{tsakalidis_predicting_2015} use a multivariate feature model, consisting of eleven features from tweets and one from the polls, from three different European countries such as Greece, Netherlands, and Germany to predict the election results. The authors use data only for seven days, starting from nine days until two days before the elections. The authors use the Weka tool for their proposed features models, in order to use the off the shelf algorithms. Results show that the choice of features outperforms the earlier works in MAE.

A study by Eom et al.~\cite{eom_Twitter-based_2015} correlates the tweet volumes with political attention based on data from two different countries: Italy and Bulgaria during the general elections in 2013 in both countries and the European Parliament elections of 2014 in Italy. The authors use stochastic and geometric Brownian equations to find the tweet volumes. The study shows that tweet volume has fluctuations, attributed to the context, and can systematically be mapped from the mass media, leading to the unreliability in predicting the election outcomes. However, in shorter periods of time, tweet volumes can have a positive correlation with the election outcomes.

Prasetyo and Hauff~\cite{dwi_prasetyo_Twitter-based_2015} use Twitter data to predict the 2014 election results in Indonesia. This work shows that predictions based on Twitter data outperform the offline poll results. The proposed scheme also suggests removing spam accounts and bias in the collected data. Other features that affect the prediction are gender, location, keywords, sentiments, and time period before the election for which the tweets are collected. To study the effect of bias in social media and polls in election prediction, Anuta et al.~\cite{anuta_election_2017} use a sentiment analysis approach where bias can be either in favor of or against the candidate. Also, the authors highlight the issue of data intentionally skewed by the source to introduce bias. The study shows, based on the US 2016 election, that Twitter and other media had a bias against both Donald Trump and Hilary Clinton. Such bias can actually affect the result prediction solely made on social media data. A study by Sanders et al.~\cite{sanders_using_2016} predicts the results of two Netherlands' elections in 2012 and 2015, respectively. The authors show the effect of demographics on the prediction. Both APIs and human annotators are used to get the gender and age of Twitter users. This study shows that additional demographic-based approach results better than just counting the mention of political parties. However, the prediction on the 2012 election has a lower MAE than 2015 election predictions. 
Cameron et al.~\cite{RePEc:wai:econwp:13/08} show that the size of the social media network has a statistically important correlation with election outcomes, based on Facebook and Twitter data. It uses regression models on social media for the New Zealand election in 2011. Although the effect of such a correlation is rather small,  it can have an effect on the results of the election with the close contest where the 'win-margin' is not so large. The authors identify several reasons for such a phenomenon. Among others reasons, the quality of the link between followers and political identities is important, as the \textit{cost} of forming such a connection is quite low and is different on both network such as users can easily follow any politician. 

Burnap et al.~\cite{burnap_140_2016} use Twitter data from the UK elections in 2015 to predict the seats for political parties. The methodology considers the positive sentiment as the voting intention for the political party. Proposed method sums up all the positive sentiment scores for a party and compares the scale of positive sentiments among all parties. It also considers the results of the previous election as given seat predictions are adjusted accordingly. Metaxas et al.~\cite{metaxas_how_2011} provide guidelines on election predictions and highlight the design of a scientific method to predict the election based on data that is systematically collected considering the important factors such as removing spam and propaganda in the data.  A study by Bhattacharya et al.~\cite{bhattacharya2016perceptions} uses the Adjective Check List (ACL) method to find people’s opinion on politicians based on different personality traits and the change in these traits over the time period. Indeed, social networks features used in most of the works are not consistent for longer duration and can lead to a random result. Issenberg~\cite{issenberg2012obama} also highlights the rigorous use of votes data in Obama’s Campaign 2008.

This section covered the work related to the use of social media in political campaigns and to predict elections results. Most of the work mentioned here use social media data, mainly Twitter, that is collected around the election times with respect to the country being studied. Most common methods consist of finding sentiments of users for political parties, politicians, and topics along with studying tweet publishing statistics. 

\begin{table}
\caption{Studies Focused on Politics and Election Predictions}
\begin{center}
\begin{tabular}
{p{0.30\linewidth}p{0.15\linewidth}p{0.15\linewidth}p{0.28\linewidth}}
\toprule
\multicolumn{1}{c}{\textbf{Paper}} & \multicolumn{2}{c}{\textbf{Feature\footnote{`*' shows the feature used in method. No `*' sign in any of the columns shows that work is a qualitative study based on survey or interviews} }}  &\multicolumn{1}{c}{\textbf{Method}} \\
\toprule

  & \textbf{Linguistics} & \textbf{ Network} & \\
\cmidrule{1-4}

Ruby Siegel. 2018 \cite{siegel_life_2018}  & 	&&	Survey, Statistical analysis \\
\hline
Hoffman et al. 2017\cite{hoffmann2017facebook} &  	&&	Quantitative Survey,Linear Regression\\

\hline
Semaan et al. 2014\cite{semaan2014social}  & 	&&	Survey, Qualitative Study\\
\hline
Agarwal et al. 2019\cite{agarwal2019tweeting} &*&*& Statistical Analysis , LIWC\\
\hline
Kulshrestha et al. 2017 \cite{kulshrestha2017politically}  & 	*&*&	Information Diffusion, augmented contagion model\\
\hline
Li et al. 2018\cite{li2018working} & 	*&*&	LIWC, Statistical Measures, Empowerment Theory\\
\hline
Khairun-nisa Hassanali and Vasileios Hatzivassiloglou. 2010 \cite{hassanali2010automatic} & *	&&	SVM, Named Entity Recognition\\
\hline
Oliveira et al. 2017\cite{oliveira2017can} & 	*&&	sentimental analysis, comparison b/w social media and survey polls\\
\hline
Tumasjan et al. 2010 \cite{tumasjan_predicting_2010} & *	&&	LIWC 12 dimensions\\
\hline
Kagan et al. 2015\cite{kagan_using_2015} & 	*&&	sentiment analaysis\\
\hline
 Dokoohaki et al. 2015\cite{dokoohaki_predicting_2015} & *	&*&	link prediction, strength\\
\hline
Tsakalidis et al. 2015\cite{tsakalidis_predicting_2015} & *	&&	multivariate feature model\\
\hline
Eom et al. 2015\cite{eom_Twitter-based_2015} & *&&	Tweet Volumes,  Brownian geometric equations, Differential Equations\\
\hline
Prasetyo and Hauff. 2015\cite{dwi_prasetyo_Twitter-based_2015} &*&&Statistical Measures\\
\hline
Anuta et al. 2017\cite{anuta_election_2017} &*&&Sentiment Analysis\\
\hline
Sanders et al. 2016\cite{sanders_using_2016} &*&&Quantitative\\
\hline
Cameron et al. 2008 \cite{RePEc:wai:econwp:13/08} &&*&Regression Analysis\\
\hline
Burnap et al 2016\cite{burnap_140_2016}&*&&Sentiement Analysis\\
\hline
Metaxas et al. 2011\cite{metaxas_how_2011} &*&&Qualitative Study\\
\hline
Bhattacharya et al. 2016\cite{bhattacharya2016perceptions} &*&& Adejective Check List\\

\bottomrule

\end{tabular}
\label{t_election_predictions}
\end{center}
\end{table}


\subsection{System Design} \label{suubsec:exploratory}
In this section we present work that proposes a full system design revolving around some special research related to computational politics. 

Cambre et al.~\cite{cambre_escaping_2017} propose a system design that can help to break the echo chamber effect and moderate the online political discussion. They propose a variant of the Talkabout~\cite{kulkarni2015talkabout} online system for this purpose that could lead to more ideologically and politically diverse discussions. 

Most of the online communication tools are not specifically designed for information seeking and dissemination, especially in the context of social media. Semaan et al.~\cite{semaan_designing_2015} use  user surveys to find the shortcomings in present media for information seeking and sharing from participants and proposes a system called \textit{Poli} to address the problems highlighted by the users.  
Kleinberg and Mishra~\cite{kleinberg_psst:_2008} find the political discourse of the politicians and how consistent they are in their opinions. They develop a search engine based on the keyword extraction from the different sources and linking them to the politicians. 

Robertson et al.~\cite{dade-robertson_political_2012} discuss the use of situational media to help the decision-makers and politicians to get instant feedback on important issues. This study introduces a sensory device like a mobile voting system that politicians can deploy during important political events (e.g. elections, protests) to get people's opinions.
A study by Liu et al~\cite{liu_scalable_2015} presents a salable computational system to model and map the political redistriction problem.  Meanwhile, Rill et al.~\cite{rill2014politwi} develop a system called "PoliTwi" to identify emerging top topics earlier than Google Trends, based on tweets in 2013 parliamentary elections. The system uses a pipeline of different steps ranging from a crawler to pre-processing steps including hashtags and sentiment analysis. This processed information is fed to the analysis module that uses statistical measures based on the Gaussian distribution of the topics to predict the topics and display them on the mobile and web interfaces of the system. Awadallah et al.~\cite{awadallah_opinionetit:_2011} propose a system called 'Opinionetit' to develop a knowledge base for the people and their opinion about politically controversial topics. The authors develop rules using a lexicon originating from both supporting and opposing opinions on given facts. It achieves 72\% precision. 
Finally, Patwari et al.~\cite{patwari_tathya:_2017} give an annotated dataset with fact checks and present a system called TATHYA to predict the fact-checking statements from political documents automatically. The system is based on the linguistic features of political debates and achieves a 0.21 F1 score.

This section focused on the studies where many researchers have proposed and developed a system for tasks related to political opinion and information checking. Much the same as the prior section, methods for opinion and fact checking are based on the linguistics based approaches.

\begin{table}
\caption{ Studies Focused on System Designs}
\begin{center}
\begin{tabular}
{p{0.30\linewidth}p{0.28\linewidth}}
\toprule
\multicolumn{1}{c}{\textbf{Paper}} &  \multicolumn{1}{c}{\textbf{System }} \\
\toprule

Cambre et al. 2017 \cite{cambre_escaping_2017} & Discussion System based on diversification \\
\hline
Semaan et al. 2015 ~\cite{semaan_designing_2015} & Information Curation and Sharing system \\
 \hline

 Kleinberg and Mishra. 2008\cite{kleinberg_psst:_2008} & Search Engine \\
 \hline

 Robertson et al. 2012\cite{dade-robertson_political_2012}  & Situational Media \\
 \hline

Liu et al. 2015 \cite{liu_scalable_2015} &Political Redistribution  \\
 \hline

 Rill et al. 2014\cite{rill2014politwi} & Trending Topics Detection \\
\hline

Awadallah et al. 2011 \cite{awadallah_opinionetit:_2011} & Opinion Monitoring \\
 \hline

Patwari et al. 2017 \cite{patwari_tathya:_2017} & Fact Checking  \\

\bottomrule

\end{tabular}
\label{t_system_design}
\end{center}
\end{table}

\section{Discussion and Future Directions} \label{sec:discussion}

We observe from the above categorization and survey of the work that computational politics covers a broad range of topics and provide an enormous landscape upon which to build. Network structures and linguistic characteristics of the data are the primary sources to find answers to the computational politics problem domain. In this survey, we primarily mentioned works that achieve reasonably good results. However, we have to also report important limitations in these areas. The foremost limitation for most of the works is the reliability of the results and their generalization. For instance, election predictions need to be verified over more than a single election. It is also important to mention that this limitation is not limited to computational methods and is mainly dependent on data. The political affiliation studies we report on mainly rely on very limited annotated data sets. It can affect the accuracy of the algorithms. Most computational approaches rely on sentiment analysis and topic modeling. However, such sentiments can be affected by contextual factors that are not present in the data features used for training classifiers. Research ethics and data privacy are another challenge. Research in Computational Politics can be used for unfair competition. We have seen the recent example of Cambridge Analytica~\cite{confessore2018cambridge}. There is a need for policymakers and researchers to define the regulations for availability and the transparent and fair use of data.

For electoral predictions, most studies use the data from a low number of days and also lack the real-life testing of results. Such studies rely on a lower set of analysis, and there is a lack of methods to correlate the contextual information and demographic complexities. Another limitation in this area is a lack of focus for individual constituencies or prediction at lower levels rather than for a party as a whole. We further highlight future research directions based on the challenges and limitations of the studies.






\subsection{Data Interoperability and Contextual Information}
The heterogeneity of data sources and the dynamic nature of users creates unreliability of results and reduces the reliability of the computational methods.

\subsubsection{Annotated Dataset}
Researchers should also focus on developing and sharing the datasets of annotated users and other entities such as media outlets. The availability of such data will make comparative studies easier, which in turn will help in the development of more reliable algorithms. Based on the studies reported here, the time period during which data is collected has a significant importance, e.g., the data collected right before an election may give more accurate results for election predictions. However, more longitudinal studies should be performed on the same users for longer time periods to effectively model and present user behaviors in politics. 

\subsubsection{Active Learning Methods for Data Annotation} The limitation of annotated datasets and the amount of resources required to annotate datasets is another challenge. New computational methods for the active learning of data should be explored where more data can be annotated with less human effort. 

\subsubsection{Knowledge Graphs} A future direction in this area is the development of systems which can semantically relate the data from different sources but can also relate to the context. It is important to not only focus on the textual and visible network properties but also find the latent attributes of users and networks with their temporal effects. Linguistic approaches like opinion mining and sentiment analysis can be significantly affected by the lack of contextual information as data features cannot capture all the information. Research on building knowledge graphs and connecting various sources of data can help to increase the contextual information. Furthermore, these knowledge graphs can also be used for fake news detection and fact-checking.

\subsubsection{Fast and Effective Dimensionality Reduction Methods} Such methods should be developed to effectively perform the dimensionality reduction of related data and visualize the patterns in real-time.  We have extensive temporal data, such as social media data, available but has not been effectively visualized to in this domain. Such methods will increase the efficiency of real-time predictions and analysis. 

\subsection{Unexplored Realms} 
Most of the work that we mention here focuses on data collected from developed countries, except for a few other developing countries, having a particularly stronger democratic system. There is a large user base in terms of social networks in less explored countries. However, one reason for such a lack of focus is the language processing barrier. 
It creates two-fold challenges for the development of the lexicons and computational linguistic methods for the languages of those countries. Exploratory studies can also help social scientists to reason about the evolving political systems in developing countries. 

Authoritarian regimes are also among the least explored areas. Researchers can focus on data from those countries and political discourse from that data can give useful insights into socio-political issues. 
Localized Political Lexicons need to be developed for both the less explored languages and systems. There are more than one democratic and electoral processes around the worlds with certain terminologies. Apart from differences in the process, there are language differences as well. There are lexicons available for sentiment detection, but there are no robust and generalizable lexicons for the political classification of text. Future research directions may include developing lexicons for political purposes. 




\subsection{Governance and Social Media}

Internet is a medium with the ability to deliver messages virtually everywhere in the world. However, different governments control such freedom of expression and thus, the reach of the content. In this survey, we have discussed studies on how social media led campaigns results in government changes or huge uprisings. This leads stakeholders to develop software and tools that can detect in advance and monitor such activities in real-time. On the other hand, social media can be used to help organizations for better governance. We have seen organizations taking note of viral content on the Internet. Such systems can be designed to highlight the public sentiments and opinions surrounding the issues and government policies where political parties can use real-time opinion/sentiment mining systems. Grossklags et al.~\cite{grossklags2011young} highlight that American users did not find the social referral links on government websites to be an effective medium to contact the organizations. Merely providing the contacts and referral does not result in better engagement. Participatory design specialists should explore  ways for administrators and common people to easily communicate with each other.


Another aspect of governance comes with privacy laws. Protection of users' privacy and rights is a major concern while dealing with personal information and particularly when this information can be used to disclose latent features of users. Researchers and other stakeholders try to take the best from the data. It is therefore crucial to design, develop, and practice such ethics and privacy policies which can help to protect the users' rights on the information they share. Companies should not be able to use information data without prior information and consent. 


\subsection{From Social Media to Consensus Formation Tools}

Doris-Down et al.~\cite{doris-down_political_2013} show that echo chambers can lead to more polarized opinions and increased interaction from different partisan can help overcome the echo chamber effect. To show results, they develop a mobile application named "Political Blend" giving people a platform to allow people meet in real life and discuss political issues. It can allow two strangers with opposite views to interact with each other. 
The current design of social networks and online communities (data sources for computational politics) revolves around the idea of personalization. Such personalization leads to a more self-centered approach, the censorship of opposing views, and the buildup of echo chambers. Such systems designs should be studied where personalization should not be at the cost of other useful factors such as constructive engagement and discussions. From a computational perspective, these design methods should be investigated where information can easily cross the bubble effect around users. One aspect of such a design is the consensus. The idea of consensus formation tools~\cite{reviewer1976scenario}  is not new. However, it gives another dimension in Internet based mediums for tasks related to our problem domain. It can lead to less polarization of social media and better recommendations of the content. 

Constructive engagement along with contextual factors may bring change in political affiliations and voting tendencies from one party to another. People may change their opinions on certain topics and personalities from time to time. Such changes can be reflected in people's online interaction patterns. However, a small number of studies focus on changes of political affiliation, opinions or finding the change point in the temporal analysis. A combination of long-term data with contextual information will lead to a better study of opinion discourse.







\section{Conclusion}\label{sec:conclusion}

In this paper, we described the domain of computational politics and how traditionally data has been used for political science-related tasks. We summarized for the first time the related work in one place and present the categorization of computational politics in five major areas: (1) Community and user modeling; (2) Information flow; (3) Political discourse; (4) Election campaigns; and (5) System Design. We also presented the descriptive frameworks for two primary data sources: social media and political debate. We categorized these frameworks from the user and data perspective. We presented prominent works in each category to highlight the recent trends. Furthermore, we reported the limitations and challenges of computational approaches, followed by possible future research directions. We recommend the design of systems where partisan echo chambers do not compromise on the political discussion of different opinions and have less bias in information flow. We also recommend the political data solicitation and analysis system which can give an understanding beyond the linguistic and platform-specific limitations, be generalizable and have the contextual embedding in it.

\bibliographystyle{ACM-Reference-Format}
\bibliography{acmsmall-sample-bibfile,survey,affiliation,miscel,bias}










\end{document}